\documentclass[sip,biber]{now-journal} 
\usepackage[utf8]{inputenc}
\usepackage{amssymb}
\usepackage{float}
\usepackage{amsmath,amsfonts}
\usepackage{algorithmic}
\usepackage{algorithm}
\usepackage{array}
\usepackage[caption=false,font=normalsize,labelfont=sf,textfont=sf]{subfig}
\usepackage{threeparttable}
\usepackage{textcomp}
\usepackage{stfloats}
\usepackage{url}
\usepackage{verbatim}
\usepackage{graphicx}

\title{Serial-OE: Anomalous sound detection based on serial method with outlier exposure capable of using small amounts of anomalous data for training}

\author[1*]{Ibuki Kuroyanagi}
\author[1]{Tomoki Hayashi}
\author[2]{Kazuya Takeda}
\author[3]{Tomoki Toda}

\affil[1]{The Graduate School of Informatics, Nagoya University, Aichi, Japan}
\affil[2]{The Institutes of Innovation for Future Society, Nagoya University, Aichi, Japan}
\affil[3]{The Information Technology Center, Nagoya University, Aichi, Japan}

\defbibheading{bibliography}[\refname]{}

\DeclareSourcemap{
	\maps[datatype=bibtex, overwrite=true]{
		\map{
            \step[fieldsource=booktitle,
    			match=\regexp{.*CVPR.*},
    			replace={Proc. Computer Vision and Pattern Recognition}]
            \step[fieldsource=booktitle,
    			match=\regexp{.*ICCV.*},
    			replace={Proc. International Conference on Computer Vision}]
           \step[fieldsource=booktitle,
    			match=\regexp{.*IJCNN.*},
    			replace={Proc. International Joint Conference on Neural Networks}]
		    \step[fieldsource=booktitle,
			match=\regexp{.*EUSIPCO.*},
			replace={Proc. European Signal Processing Conference}]
			\step[fieldsource=booktitle,
			match=\regexp{.*DCASE2019.*},
			replace={Proc. Detection and Classification of Acoustic Scenes and Events 2019 Workshop}]
			\step[fieldsource=booktitle,
			match=\regexp{.*DCASE2020.*},
			replace={Proc. Detection and Classification of Acoustic Scenes and Events 2020 Workshop}]
			\step[fieldsource=booktitle,
			match=\regexp{.*DCASE2021.*},
			replace={Proc. Detection and Classification of Acoustic Scenes and Events 2021 Workshop}]
			\step[fieldsource=booktitle,
			match=\regexp{.*DCASE2022.*},
			replace={Proc. Detection and Classification of Acoustic Scenes and Events 2022 Workshop}]
            \step[fieldsource=booktitle,
			match=\regexp{.*DCASE2023.*},
			replace={Proc. Detection and Classification of Acoustic Scenes and Events 2023 Workshop}]
            \step[fieldsource=booktitle,
			match=\regexp{.*DCASE2024.*},
			replace={Proc. Detection and Classification of Acoustic Scenes and Events 2024 Workshop}]
			\step[fieldsource=booktitle,
			match=\regexp{.*ICPICS.*},
			replace={Proc. International Conference on Power, Intelligent Computing and System}]
			\step[fieldsource=booktitle,
			match=\regexp{.*Computer.*Vision.*and.*Pattern.*Recognition.*},
			replace={Proc. Computer Vision and Pattern Recognition}]
			\step[fieldsource=booktitle,
			match=\regexp{.*Interspeech.*},
			replace={Proc. Interspeech}]
			\step[fieldsource=booktitle,
			match=\regexp{.*ICASSP.*},
			replace={Proc. International Conference on Acoustics, Speech and Signal Processing}]
			\step[fieldsource=booktitle,
			match=\regexp{.*International.*Conference.*on.*Learning.*Representations.*},
			replace={Proc. International Conference on Learning Representations}]
			\step[fieldsource=booktitle,
			match=\regexp{.*International.*Conference.*on.*Machine.*Learning.*},
			replace={Proc. International Conference on Machine Learning}]
			\step[fieldsource=booktitle,
			match=\regexp{.*ISIE.*},
			replace={Proc. International Symposium on Industrial Electronics}]
			\step[fieldsource=booktitle,
			match=\regexp{.*Automatic.*Speech.*Recognition.*and.*Understanding.*},
			replace={Proc. ASRU}]
			\step[fieldsource=booktitle,
			match=\regexp{.*Spoken.*Language.*Technology.*},
			replace={Proc. SLT}]
			\step[fieldsource=booktitle,
			match=\regexp{.*Speech.*Synthesis.*Workshop.*},
			replace={Proc. SSW}]
			\step[fieldsource=booktitle,
			match=\regexp{.*workshop.*on.*speech.*synthesis.*},
			replace={Proc. SSW}]
			\step[fieldsource=booktitle,
			match=\regexp{.*Advances.*in.*neural.*information.*processing.*},
			replace={Proc. Neural Information Processing Systems}]
			\step[fieldsource=booktitle,
			match=\regexp{.*Advances.*in.*Neural.*Information.*Processing.*},
			replace={Proc. Neural Information Processing Systems}]
			\step[fieldsource=booktitle,
			match=\regexp{.*Workshop.*on.* Applications.* of.* Signal.*Processing.*to.*Audio.*and.*Acoustics.*},
			replace={Proc. Workshop on Applications of Signal Processing to Audio and Acoustics}]
			\step[fieldsource=series,
			match=\regexp{.+},
			replace={{}}]
			\step[fieldsource=editor,
			match=\regexp{.+},
			replace={{}}]
			\step[fieldsource=publisher,
			match=\regexp{.+},
			replace={{}}]
			\step[fieldsource=month,
			match=\regexp{.+},
			replace={{}}]
			\step[fieldsource=location,
			match=\regexp{.+},
			replace={{}}]
			\step[fieldsource=address,
			match=\regexp{.+},
			replace={{}}]
			\step[fieldsource=organization,
			match=\regexp{.+},
			replace={{}}]
		}
	}
}

\AtEveryBibitem{
  \clearfield{doi}
  \clearfield{url}
}

\issuevolumeyear{2024}
\issuevolumenumber{xx}
\issueelocation{e} 
\articletype{Original Paper} 

\articledatabox{Received xx xxxxx 2024; Revised xx xxxxx 2024\\[2pt]
ISSN 2048-7703; DOI 10.1561/116.xxxxxxxx\\
\copyright\ 2024 I. Kuroyanagi, T. Hayashi, K. Takeda and T. Toda}

\OAstatement{This is an Open Access article, distributed under the terms of the Creative Commons Attribution licence (\url{http://creativecommons.org/licenses/by-nc/4.0/}), which permits unrestricted re-use, distribution, and reproduction in any medium, for non-commercial use, provided the original work is properly cited.}

\keywords{Anomalous sound detection, Self-supervised learning, Outlier exposure, Serial method.}

\creditline*{Corresponding author: Ibuki Kuroyanagi, kuroyanagi.ibuki@g.sp.m.is.nagoya-u.ac.jp.
This paper was partly supported by Japan’s New Energy and Industrial Technology Development Organization (NEDO), project JPNP20006.
}

\begin{document}

\begin{abstract}
We introduce Serial-OE, a new approach to anomalous sound detection (ASD) that leverages small amounts of anomalous data to improve the performance.
Conventional ASD methods rely primarily on the modeling of normal data, due to the cost of collecting anomalous data from various possible types of equipment breakdowns.
Our method improves upon existing ASD systems by implementing an outlier exposure framework that utilizes normal and pseudo-anomalous data for training, with the capability to also use small amounts of real anomalous data.
A comprehensive evaluation using the DCASE2020~Task2 dataset shows that our method outperforms state-of-the-art ASD models. 
We also investigate the impact on performance of using a small amount of anomalous data during training, of using data without machine ID information, and of using contaminated training data.
Our experimental results reveal the potential of using a very limited amount of anomalous data during training to address the limitations of existing methods using only normal data for training due to the scarcity of anomalous data.
This study contributes to the field by presenting a method that can be dynamically adapted to include anomalous data during the operational phase of an ASD system, paving the way for more accurate ASD.
\end{abstract}

\section{Introduction}
Anomalous sound detection (ASD) is used to detect unusual sounds, such as equipment malfunctions.
It is often used to monitor sound inside machines, or other areas that cannot be directly monitored by cameras, such as equipment in tunnels or other dark locations.
Early detection of anomalous conditions can minimize losses due to outages and damage resulting from breakdowns. Due to a decrease in the number of skilled maintenance technicians in recent years, and an increase in the number of aging facilities~\cite{Chen2018}, there is a greater need for systems that can automatically detect anomalous conditions and allow timely repair after mechanical breakdowns~\cite{Koizumi_DCASE2020_01,ad_survey2009,hayashi2018anomalous,Kawaguchi2019icassp}.

First, let’s consider the development of an ASD system. 
Ideally, both normal and anomalous audio data would be used to train a binary classification model.
However, the possible types of anomalous data are diverse, and intentionally causing machines to break down in various ways to collect anomalous data is expensive. 
Therefore, the detection model of an ASD system is often trained using only normal data~\cite{Koizumi2017OptimizingAF,Koizumi2019UnsupervisedDO}. 
Next, let’s consider the operational phase of ASD systems.
These systems are generally installed in facilities that experience occasional breakdowns due to aging machinery, or because the facility’s design makes human monitoring impractical.
Conversely, facilities that do not generally experience anomalous conditions do not need an ASD system.
Therefore, facilities that have installed ASD systems are likely able to capture samples of various types of anomalous audio data during breakdowns, data which contains important information about possible anomalous conditions for the machine, so it would be advantageous to utilize this data to improve the performance of the ASD system. 
Several methods have been proposed using active learning to improve performance in sound event detection tasks by utilising a small number of labelled data~\cite{Shuyang2018,Meire2023}.
However, most studies on ASD have focused on the development phase, when normal data is generally used to create the detection model for the ASD system. 
Few studies have focused on continuously improving ASD performance by utilizing the small amounts of anomalous data that can be obtained during the operational phase~\cite{kuroyanagi2021anomalous}.

Current ASD methods can be classified into generative and discriminative model-based methods~\cite{Koizumi_DCASE2020_01,ad_survey2009}. 
Generative model-based methods identify sounds deviating from the normal distribution as anomalous data.
For example, some methods use reconstruction errors obtained using an autoencoder~\cite{uefusa2020,Giri2020b,Purohit2020,Kapka2020,Mishra2021}, Long-Short Term Memory networks~\cite{muller2021a}, WaveNet~\cite{hayashi2018anomalous}, or generative adversarial networks  ~\cite{XIA2022497,JiangAnbai2023}, as well as methods that use the likelihoods of Gaussian mixture models (GMM)~\cite{Reynolds2009,liu2019,GuanJian2023} or normalizing flow~\cite{NIPS2017_6c1da886,NEURIPS2020_NF,Dohi2021icassp}.
These generative model-based methods cannot use anomalous data because they are trained while generating normal data. 
In methods that minimize the reconstruction error of normal data, if anomalous data is added to the training data and the reconstruction error is minimized, the anomalous data can be reconstructed, so the ASD performance is degraded. 
Similarly, if anomalous data is added to methods that maximize the likelihood of normal data, the likelihood of anomalous data increases, so the performance is degraded. In order to utilize anomalous data with these methods, it is necessary to add an architecture for utilizing anomalous data to the original method.

Discriminative model-based methods can also be divided into two types.
The first method uses normal data to make multi-class classifications.
It includes methods based on classification by machine type, using either data augmentation~\cite{Inoue2020}, metadata classification~\cite{Giri2020a,stgrammfn2022,Wilkinghoff2021a,Wilkinghoff2021b,Wilkinghoff2023,Chen2023wavenet,Wilkinghoff2024}, or contrastive learning~\cite{Hojjati2022,geco2023,Guan2023,choi2024noisy}, which have demonstrated competitive performance in recent years.
These methods compute anomaly scores for multiple types of machines, and have the advantage of being able to detect anomalous conditions in multiple machines using a single model.
However, because these methods do not assume the use of anomalous data, it is not easy to use them without changing the architecture, even if anomalous data is available.
The second method defines normal data and pseudo-anomalous data and use a binary classification.
Some methods use a single model to infer the status of a single machine~\cite{kuroyanagi2021anomalous,ocsvm1999,ruff2018deep,Ruff2020Deep,primus2020anomalous,Kuroyanagi2021dcasew}, while others use a single model to infer the status of multiple machines by considering the normal sounds of all similar machines as normal data~\cite{kuroyanagi2022eusipco}. 
Since these methods use pseudo-anomalous data as training data, it is possible to utilize real anomalous data without making any changes to the architecture.

In this research, we also employ a discriminative model-based\\ method that defines normal data and pseudo-anomalous data and use a binary classification, allowing us to take advantage of any real-world anomalous data that is available as additional training data, in order to improve ASD performance.
In our previous work, we compared the average performance of the proposed method when using the DCASE2021 Task2 dataset to show that it improves ASD performance~\cite{kuroyanagi2022eusipco}.
However, we did not perform a component-by-component analysis of our method, evaluate the effectiveness of using of real-world anomalous data for training, or examine the possibility of contamination of the real-world anomalous data with normal data, which may occur when collecting the anomalous data. 
In this paper, we compare the performance of our proposed system with other ASD methods when using the DCASE2020 Task2 dataset, the major ASD dataset available ~\cite{Koizumi_WASPAA2019_01,Purohit_DCASE2019_01}.
Our investigation into the possibility of contamination of the real anomalous data used for model training is the first such investigation of this topic in the field of ASD.
We not only compare the proposed method with state-of-the-art methods proposed in previous studies, and with similar methods, but also perform ablation studies for each component, evaluate the effect of the DCASE2020 Task2-specific information known as ‘machine ID’, which has been used in previous studies, evaluate the effect of utilizing a small amount of real anomalous data for model training, and evaluate the effect of training the model with data containing anomalous and normal data from the same type of machine.
\footnote{The current implementation of our proposed Serial-OE system is available at \url{https://github.com/ibkuroyagi/Serial-OE}.}

The rest of this paper is organized as follows:
In Section~\ref{section:related}, we describe previously proposed ASD methods which are able to utilize anomalous data. 
In Section~\ref{section:proposed_method}, we provide the details of our proposed method.
In Section~\ref{section:exp}, we investigate the reasons why our proposed method improves ASD performance and the impact of utilizing anomalous data during detection model training on performance through experimental evaluation. 
In Section~\ref{section:limit}, we note the limitations of this study, and in Section~\ref{section:conclusion} we present our conclusions.

\begin{figure*}[!t]
\centering
\includegraphics[width=\textwidth]{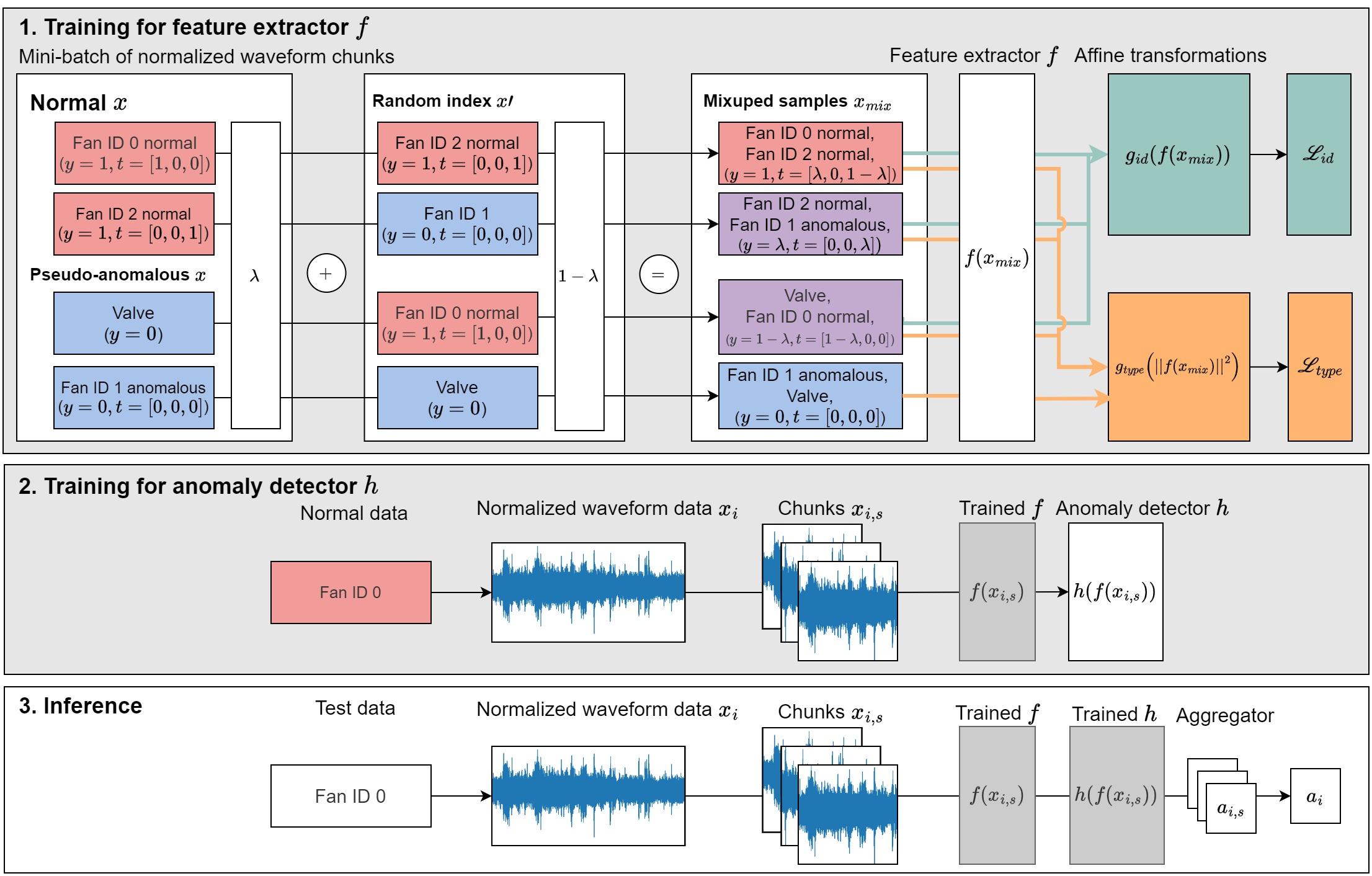}
\caption{
Overview of the proposed Serial-OE method.
The gray areas represent the training phase, while the white area represents the inference phase. 
This example shows how to train the ASD system to detect anomalies in the sound produced by machine type Fan.
Section 1 shows the training of feature extractor $f$, which is trained by applying two loss functions to the features extracted from a mixture of normal and pseudo-anomalous sounds, as sampled by a batch sampler in mini-batches.
Section 2 shows the training of anomaly detector $h$. 
Since a different anomaly detector $h$ is trained for each machine ID, here we train the anomaly detector using the features of normal audio from Fan ID 0 obtained from pretrained feature extractor $f$.
Section 3 shows the inference process used to obtain the anomaly score, which is computed by dividing the test sample into multiple chunks, each of which is fed into pretrained feature extractor $f$ and trained anomaly detector $h$, with the resulting anomaly scores aggregated to obtain the final score. 
}
\label{fig:overview}
\end{figure*}

\section{ASD methods that can utilize anomalous data}
\label{section:related}
Few methods which utilize anomalous data have been proposed for ASD.
Here, small amounts of anomalous data refers to data that accounts for approximately 0.1\% to 3\% of the training data.
DDCSAD+BCE~\cite{kuroyanagi2021anomalous} is a method which we have proposed previously for improving the performance of ASD systems by utilizing small amounts of anomalous data for training.
A feature extractor is trained by defining normal data and pseudo-anomalous data, and two loss functions are then applied. 
The first loss function performs binary classification using binary cross-entropy.
The second loss function uses metric learning to define the centroid of each class in the feature space, then uses Euclidean distance to cluster data in the same class while pushing data from different classes farther apart.
The classification results for the normal and pseudo-anomalous classes obtained through multi-task learning using these loss functions are then used as anomaly scores. 
This method improves performance by not only discriminatively learning normal and pseudo-anomalous data, but also by explicitly classifying normal data and other data in the feature space.
This method also improves performance by classifying anomalous data as part of the pseudo-anomalous data class.

To explain this method in more detail, here we will describe the DCASE2020 Task2 data set~\cite{Koizumi_WASPAA2019_01,Purohit_DCASE2019_01} and how it was used when evaluating this method.
The Task2 dataset has a hierarchical structure of six machine types (Fan, Pump, Slider, Valve, ToyCar, and ToyConveyor), as well as machine IDs that identify the manufacturer and model of each machine belonging to each machine type.
There are seven machine IDs for each machine type, except for the ToyConveyor machine type, which contains only six machine IDs, thus there are 41 machine IDs in the entire data set.
DDCSAD+BCE treats specific machine IDs within a machine type as normal sound data, and the normal sounds of the other machines of the same type, as well the normal sounds of the other types of machines, are used as pseudo-anomalous data.
For example, if machine ID 0 of machine type Fan is considered to be our normal data, the normal sounds of the Fans with machine IDs from 1 to 6, as well as the normal sounds of the different machine types, such as Pumps,  Sliders, Valves, ToyCars and ToyConveyors are treated as pseudo-anomalous data.
Therefore, 41 different models need to be trained when using the DCASE2020 Task2 data set in order to evaluate ASD performance for all of the machines.

Our DDCSAD+BCE method performed well in terms of average AUC for a single model at that time. Furthermore, ASD performance was improved by adding real anomalous data to the pseudo-anomalous class and by fine tuning the detection model.
However, the performance of this method is not as high as that of more recent ASD methods, and it also suffers from large variations in performance from machine ID to machine ID.
Since the proposed method trains a feature extractor for each machine ID, performance is strongly influenced by the particular characteristics of each machine’s audio data, which is defined as normal data.
In addition, this method uses only the classification probability from discriminative training as the anomaly score, however discriminative classification results are not necessarily equivalent to ASD results~\cite{stgrammfn2022}.
Considering the recent success achieved by combining discriminative and generative models, our method’s lack of an anomaly detection method based on generative models, such as outlier detection, may also be a factor in its lower performance~\cite{Kawaguchi2021}. 
Therefore, to take advantage of the small amount of anomalous data, this paper retains the framework for classifying normal and pseudo-anomalous data but combines ASD methods based on generative models to improve performance.

From a different perspective, we investigated how anomalous detection in image detection using anomalous data can be used to improve the performance of ASD systems.
Random images obtained from large data sets are called outlier exposures, and the use of such data has resulted in higher anomaly detection performance~\cite{hendrycks2018deep,Hendrycks2019}.
DeepSAD~\cite{Ruff2020Deep} was the first deep model to exploit a small number of anomalous cases, by generalizing unsupervised DeepSVDD~\cite{ ruff2018deep} for use in a semi-supervised anomaly detection setting.
In~\cite{ruff2020rethinking}, Ruff $et~ al.$ further modified DeepSAD using a method based on cross-entropy classification, significantly improved DeepSAD’s performance.
However, methods that use only a small amount of anomalous data have a problem; their output is biased against the known anomalous data, resulting in poorer performance.
DAR~\cite{ding2022catching} and BGDA~\cite{Yao2023} have been proposed to solve this problem.
By training these methods to acquire a feature space that explicitly distinguishes normal data from the small amount of anomalous data obtained, they improve the detection performance of the small amount of known anomalous data used during training without reducing the detection performance of unknown anomalous data not used during training. 
This paper also improves the detection performance for known anomalous data by explicitly classifying a small amount of anomalous data in the feature space.

These methods have achieved good performance but they are difficult to apply directly to ASD because many of the techniques used are unique to image data processing.
These methods utilize CutMix~\cite{cutmix_Yun,ding2022catching}, image rotation~\cite{sohn2021learning} and semantic segmentation~\cite{Yao2023} in order to generate pseudo-anomalous data for training.
The nature of the data used to represent images and sounds differs greatly, so many factors must be considered when applying these outlier exposure methods to ASD.
In particular, sound and images differ in their approach to adding data.
For example, images generated by superimposing different images on top of each are not common in the real world and thus have a high probability of being anomalous.
However, sounds generated by the superimposing different sounds may be normal ($e.g.$, adding background noise to a movie scene). 
Therefore, it is necessary to carefully design the training process with these points in mind.

\begin{figure}[!t]
\centering
\resizebox{0.5\columnwidth}{!}{
\includegraphics[width=0.8\columnwidth]{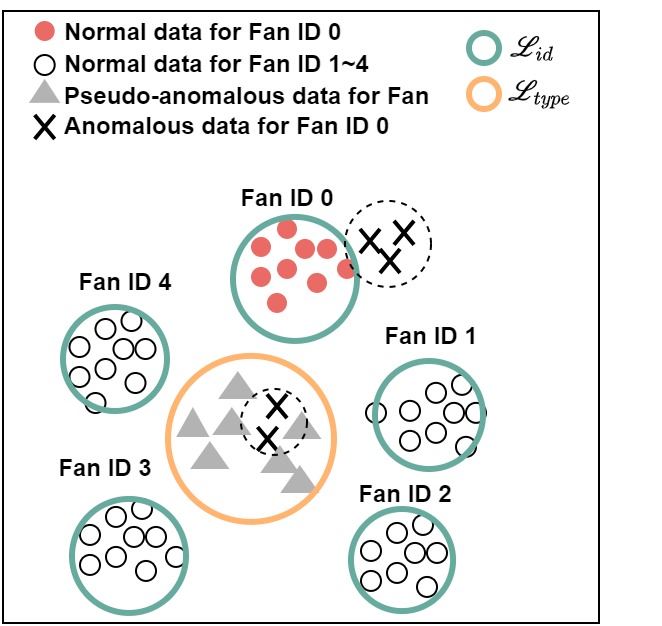}
}
\caption{
 Feature space obtained when using a feature extractor for machine type ‘Fan’. 
 The symbols $\circ$ and $\bigtriangleup$ represent the normal and pseudo-anomalous data used for training, respectively.
 The pseudo-anomalous data are distributed near the origin ($i.e.$, center) of the hypersphere after performing binary classification based on the norm of $\mathcal{L}_\mathrm{type}$, while the normal data are distributed farther from the origin.
 Data for each machine ID is distributed into separate clusters using $\mathcal{L}_\mathrm{id}$.
 For the purpose of illustrating our process when using the obtained feature space, normal data for Fan ID 0 are represented with a red $\circ$, while anomalous data for Fan ID 0 are denoted by $\times$.
 The black dashed circles represent the areas where anomalous data are distributed.
 It is assumed that anomalous data for Fan ID 0 will cluster around the normal data in the feature space if its characteristics are similar to the normal data ($e.g$. if anomalous data is caused by a slight scratch), and around the pseudo-anomalous data cluster if its characteristics significantly differ from the normal data, suggesting a possible breakdown. 
 The proposed method detects anomalous data by modeling the normal data for Fan ID 0 using a GMM, in relation to such a feature space.
}
\label{fig:feature_embed}
\end{figure}

\section{Proposed method}
\label{section:proposed_method}
\subsection{Overview}
When designing the proposed method, we decided that if it was too difficult to collect anomalous audio data for the target device during the development of an ASD system, the anomaly detection model could be trained using only normal data. 
But if it was possible to collect a small amount of anomalous data for the target device, the system would allow the use of that data to improve anomaly detection performance. 
With this in mind, we developed Serial-OE, which is an ASD system consisting of a DNN-based feature extractor and a GMM-based anomaly detector in a serial method~\cite{kuroyanagi2022eusipco}.
This framework incorporates the powerful feature extraction capability of DNNs for high-dimensional data, with the high detection performance of generative models, allowing Serial-OE to achieve performance comparable to that of other recently proposed ASD methods~\cite{muller2021a,Wilkinghoff2021a,Kawaguchi2021,sohn2021learning,moritaSECOM2021,Roth2022}. 
The proposed method also incorporates outlier exposure as a binary classification method for identifying normal and pseudo-anomalous data.

An overview of the proposed method is shown in Fig.~\ref{fig:overview}. 
Incorporating outlier exposure in our proposed method makes it easier to utilize anomalous data when it is available. 
The proposed method trains one feature extractor DNN for each machine type, and one GMM for each machine ID. 
Therefore, in the case of the DCASE2020 Task2, there are six machine types, so six feature extractors are trained, and 41 GMMs are trained for use as anomaly detectors since there are 41 machine IDs. 
Each sample is a mixture of machine operating and environmental sounds.
When training the feature extractors, normal and pseudo-anomalous data are used.
The normal sound of the target machine type is used as the normal data.
The normal sound of the other machine types are used as pseudo-anomalous data.
When real anomalous data for a machine is available, this anomalous data is treated in the same way as pseudo-anomalous data.
Note that the proposed method assumes that both pseudo-anomalous data and anomalous data are chunks of the same length as normal data.
When training the anomaly detectors, only normal data is used.
We discuss the validity of using normal sounds from machines other than the target machine type as pseudo-anomalous data.
One of the most significant factors that degrade the performance of ASD systems is the misclassification of anomalous and environment sound noises. 
So, ASD models should be trained to ignore environmental sound noise to prevent misclassification.
A model trained by a task that identifies the operation sounds of several different machines with the same environment sound noise will ignore the effect of the environment sound noise and focus on the operation sound to classify~\cite{Wilkinghoff2024}.
Using this model in ASD reduces misclassification due to noise and improve performance~\cite{Wilkinghoff2024}.
It has also been shown experimentally that the classification of normal and pseudo-anomalous sounds by this selection improves performance~\cite{kuroyanagi2021anomalous,primus2020anomalous,kuroyanagi2022eusipco}.
Therefore, it is reasonable to select data that have different operation sounds from normal data and the same environment sound noise as normal data as pseudo-anomalous data and have them classified.

The proposed method uses the following steps to infer and obtain anomaly scores:
\begin{quote}
 \begin{itemize}
  \item {\bf Preprocessing:} The input sound is divided into $S$-second chunks, and each chunk is converted into a mel spectrogram.
  \item {\bf Feature extractor:} The mel spectrogram is input to a DNN to obtain a $D$-dimensional feature vector.
  \item {\bf Anomalous detector:} The $D$-dimensional feature vector is input to the GMM for the target machine ID to obtain the negative log-likelihood for each chunk.
  \item {\bf Aggregator:} Anomaly scores for all the chunks are obtained, and an aggregator is applied to these scores.
 \end{itemize}
\end{quote}
In the following subsections, we describe each step of the training and inference processes in detail.

\subsection{Preprocessing}
Consider $N$ samples of training data, denoted by $x_i ~|~ i = 1,2,\ldots,N$. 
Each $x_i$ is associated with a label $y_i \in \{0,1\} ~|~ i = 1,2,\ldots,N$, where 1 is normal data and 0 is pseudo-anomalous data. 
Normal data refers to the normal sounds of the target machine type, and pseudo-anomalous data refers to the normal sounds of the other machine types. 
Anomalous data for the target machine type is treated as $y_i = 0$. 
Each $x_i$ has a class representing the machine ID, denoted by $t_i \in \{1,2,\ldots, C\} ~|~ i = 1,2,\ldots,N$, where $C$ is the number of machine ID classes.
When applying the loss function, $t_i$ is converted into one-hot vector $t_{i,c} \in \{0,1\} ~|~ c=1,2,\ldots,C$.
Before inputting audio data into the model, the amplitude of $x$ is normalized by calculating the mean and variance of the amplitude across $y_i = 1$ (normal sounds).
Assuming each $x_i$ represents a sound source of $L$ seconds, these data segments are divided into chunks of $S$ seconds, after which a band-pass filter is applied, converting them into mel spectrograms. 
The resulting mel spectrogram of each chunk is denoted as $x_{i,s}$ and is input into the feature extractor.

\subsection{Loss functions for training feature extractor}
The proposed method uses two loss functions to train feature extractor $f$ to obtain a $D$ dimensional feature vector from $x_{i,s}$.
We show in Fig.~\ref{fig:feature_embed} an image of the feature space obtained by the feature extractor $f$.
In the following, we describe the training method for acquiring this feature space.
The first loss function $\mathcal{L}_{\text{type}}$ is a binary cross-entropy (BCE) that classifies data as normal or pseudo-anomalous using the norm of the feature vectors. 
Loss function $\mathcal{L}_\mathrm{type}$ is defined as follows:
\begin{equation}
\label{eq:l_type}
\begin{split}
\mathcal{L}_\mathrm{type} &=-\frac{1}{N}\sum^N_{i=1}\left\{y_i\mathrm{log}(g_\mathrm{type}(||f(x_{i,s})||^2 ))\right.\\
&\quad\left.+(1-y_i)\mathrm{log}(1-g_\mathrm{type}(||f(x_{i,s})||^2))\right\},
\end{split}
\end{equation}
where $g_{\text{type}}$ is a function that applies the sigmoid function after performing an affine transformation.
By using the norm of the feature vectors for classification, pseudo-anomalous data are distributed near the origin of the hypersphere, while normal data are distributed away from the origin. 
The loss function trains the model to classify pseudo-anomalous data and normal data with similar environmental sounds so that the norm of features for samples that do not contain operation sounds of normal data is 0. 
The model will focus on the operation sound, reducing the percentage of misclassification due to differences in environmental sound. 
In addition, by collecting environmental sound pseudo-anomalous data at a single point, the model prevents misclassification as normal data when data with a distribution that differs significantly from normal data is input.
Although anomalous data may not always be distributed near the origin at the time of inference, this loss function is introduced to reduce the risk of treating unknown data that is different from normal data as false positives.
Furthermore, adding anomalous data to the pseudo-anomalous data class will explicitly distribute data with known anomalous features near the origin.
This is expected to create a feature space that makes it easier for the anomaly detectors to detect anomalies.

The second loss function $\mathcal{L}_{\text{id}}$ is a BCE that classifies normal data by machine ID.
Note that the normal data are of the same machine type.
The loss function is defined as follows:
\begin{equation}
\label{eq:t_dash}
t'_i=y_it_{i,c},
\end{equation}
\begin{equation}
\label{eq:l_id}
\begin{split}
\mathcal{L}_\mathrm{id} &=-\frac{1}{C\sum_{i=1}^Ny_i}\sum_{i=1}^N\sum_{c=1}^C \left\{t'_i\mathrm{log}(g_\mathrm{id}(f(x_{i,s}))) \right.\\
&\quad \left.+(1-t'_i)\mathrm{log}(1-g_\mathrm{id}(f(x_{i,s})))\right\},
\end{split}
\end{equation}
where $g_{\text{id}}$ is a function that applies the sigmoid function after performing an affine transformation.
$\mathcal{L}_{\text{id}}$ is introduced with the goal of more precisely representing the characteristics of the data from the targeted machine ID, by identifying data with similar sounds.
Since the characteristics of the sounds produced by each machine ID are unique, BCE is used instead of cross-entropy.
Using BCE allows the classifier to be trained to focus on whether a sound is from a particular machine, thereby becoming more robust when mixing in other sounds. 
This will be discussed in more detail in subsection~\ref{subsection:sampling}.
It is important to note that by multiplying $t_{i,c}$ by $y_i$ in Eq.~\ref{eq:t_dash}, updates are only made for normal data. 
Thus, $\mathcal{L}_{\text{id}}$ focuses training of the feature extractor only on normal data features.

The final loss function is expressed as follows:
\begin{equation}
\label{eq:l}
\mathcal{L} = \mathcal{L}_\mathrm{type}+\alpha\mathcal{L}_\mathrm{id},
\end{equation}
where $\alpha$ is a hyperparameter to balance the loss functions. 
However, even if we train the feature extractor to update $\mathcal{L}$ as shown, stable updating is not possible due to several problems. 
Therefore, we used a few additional techniques, described in subsections~\ref{subsection:sampling} below, to stabilize the training of the feature extractor.
\subsection{Strategies for mini-batch sampling}
\label{subsection:sampling}
When training feature extractors, employing Mixup~\cite{mixupzhang2018}, as proposed in~\cite{Wilkinghoff2021a}, allows the acquisition of intermediate representations between samples, with the aim of improving performance.
However, when using the proposed method, two issues prevent the simple application of Mixup.
The first issue is the imbalance between the normal data, which consists of data from only one machine type, and the pseudo-anomalous data, which includes data from all of the machines belonging to the five other machine types.
Therefore, most of the data is pseudo-anomalous, so when Mixup is applied, Eq.~\ref{eq:l_id} is hardly updated.
The second issue is the difference in the scope of each loss function. Eq.~\ref{eq:l_type} focuses on classification of audio data into normal and pseudo-anomalous data, while Eq.~\ref{eq:l_id} focuses on the classification of normal data into machine ID classes.
Although mixing up normal and pseudo-anomalous data when using in Eq.~\ref{eq:l_type} is a normal Mixup, Eq.~\ref{eq:l_type} does not allow machine IDs to be defined for pseudo-anomalous data, so it is necessary to consider how Mixup should be used.

To address these issues, first we defined a batch sampler for creating mini-batches in order to deal with the imbalance between normal and pseudo-anomalous data.
The mini-batches created by the batch sampler contain half normal data  and half pseudo-anomalous data, so if real anomalous data can be utilized, one sample of anomalous data is selected to be included in the mini-batch, reducing the number of pseudo-anomalous samples by one.
One epoch is defined as a single update of the normal data. 
Using a batch sampler not only avoids the problem of an imbalance in data during training, but also reduces the number of iterations, making the training process more efficient.
The batch sampler also plays a role in stabilizing the training process; if training is performed without the batch sampler, the proportion of pseudo-anomalous data contained in the mini-batches will increase, and the model will continue to output values close to 0 for $\mathcal{L}_\mathrm{id}$, as the result of an unstable training process. 
In my experiment conducted without the batch sampler, $\mathcal{L}_\mathrm{id}$ diverged during the training, and anomaly scores could not be obtained. 
Therefore, it is important to adjust the balance between the ratio of normal and pseudo-anomalous data in the mini-batches.

Next, to address the difference in the scope of the two loss functions, we devised a method for mixing multiple self-supervised labels as follows:
\begin{equation}
\label{eq:x_mix}
x^{\mathrm{mix}}_i = \lambda x_i + (1-\lambda ) x_j,
\end{equation}
\begin{equation}
\label{eq:y_mix}
y^{\mathrm{mix}}_i = \lambda y_i + (1-\lambda ) y_j,
\end{equation}
\begin{equation}
\label{eq:t_mix}
t^{'\mathrm{mix}}_i = \lambda y_it_{i,c} + (1-\lambda ) y_jt_{j,c'},
\end{equation}
where $x^{\mathrm{mix}}_i, y^{\mathrm{mix}}_i, t^{'\mathrm{mix}}_i$ are samples obtained by Mixup, $x_j, y_j, t_{j,c'}$ are samples obtained by randomly shuffling the mini-batch indices, and $0 \leq \lambda \leq 1$ is the mixing coefficient, randomly sampled from Beta distribution $\mathcal{B}(\beta, \beta)$ as defined for the length of the mini-batch. 
When using Mixup, three types of samples are obtained: 
\begin{enumerate}
    \item a mixture of pseudo-anomalous data and pseudo-anomalous data, 
    \item a mixture of normal data and normal data, 
    \item a mixture of normal data and pseudo-anomalous data.
\end{enumerate}
The mixture in the second case involves data from the same or different machine IDs. 
If two normal samples $x_i$ and $x_j$ have $t_i \neq t_j$ (different IDs within the same machine type), then mixing them yields a combination of normal data from different machine IDs.
For instance, if $y_it_{i,1} = [1,0,\dots,0]$ and $y_jt_{j,2} = [0,1,\dots,0]$, then their mixture becomes $t^{'\mathrm{mix}}_i = [\lambda, 1-\lambda, 0, \dots, 0]$.
In the third case, if one sample is normal $(y_i=1)$, such as $y_it_{i,1} = [1,0,\dots,0]$, and the other is pseudo-anomalous $(y_j=0)$, such as $y_jt_{j,2} = [0,0,\dots,0]$, then their mixture yields $t^{'\mathrm{mix}}_i = [\lambda, 0,\dots,0]$.
The Mixup-generated samples $x^{\mathrm{mix}}_i, y^{\mathrm{mix}}_i, t^{'\mathrm{mix}}_i$ are applied during the training of the feature extractor by substituting them into Eqs.~\ref{eq:l_type} and~\ref{eq:l_id}.
First, let us consider Eq.~\ref{eq:l_type}, which can be updated in all cases since we only need to consider machine type $y_i$ in this loss function.
Next, consider Eq.~\ref{eq:l_id}.
A sample that is a mixture of pseudo-anomalous data and pseudo-anomalous data does not affect the update of Eq.~\ref{eq:l_id} since $y_i = 0$.
Samples containing a mix of normal and normal data or a mix of normal and pseudo-anomalous data are used to train the feature extractor to detect the proportion of normal data, since $0 \leq y^{\mathrm{mix}}_i \leq 1$.
In other words, the model is trained to focus on whether or not the sounds subject to ASD are included. 
The model is expected to detect normal data when normal sounds are included, even when the SNR of the environmental sound changes.

\subsection{Training method for anomaly detector}
A feature vector which has $D$ dimensional features is obtained from trained feature extractor $f$.
All of the audio samples of normal data for a particular machine ID in the training data are used to train the GMM functioning as anomaly detector $h$.
If $x$ represents $L$ seconds of normal data samples, the samples used to train the GMM are divided into chunks of $M = \lceil2L/S\rceil, \quad M \in \mathbb{N}$ samples, allowing for half duplication. 
Feature extractor $f$ is then applied to these chunks to obtain the features used to train the GMM.
Therefore, $M$ times more samples of data for each machine ID are used to train the GMM.
Since $M$ is assumed to be sufficiently large compared to dimensional features $D$ in this study, the full covariance matrix is used to calculate the covariance of the GMM.
\subsection{Method of aggregating anomaly scores}
During inference, each $x_i$ is divided into $M$ chunks of $S$ seconds, similar to the training process for the GMM.
As a result, $M$ anomaly score samples are obtained for each $x_i$.
ASD can detect both stationary anomalous sounds, such as the squealing of a belt or the sound of a broken propeller, and non-stationary anomalous sounds, such as a whistle or intermittent noise from a periodically operating machine. 
We explored various methods of effectively aggregating these anomaly scores. 
The ‘averaging’ aggregation method tends to highlight stationary sounds but may dilute the impact of non-stationary anomalies.
Conversely, selecting the maximum anomaly score can lead to overlooking stationary sounds, or may bias the results toward anomalies that occur in particular data segments, especially if a sound event spans multiple segments.

To mitigate these issues, we propose an intermediate aggregation strategy. 
This method involves sorting all of the chunks by their anomaly scores in descending order and then calculating the average score of those exceeding the median score.
Thus, our method of calculating anomaly scores, including our aggregation method, is as follows:
\begin{equation}
\begin{split}
&A_i = \text{sorted}\left(a_{i,1}, a_{i,2}, \ldots, a_{i,M}\right) \text{ in descending order},\\
&a_i = \frac{1}{\lceil \frac{M}{2} \rceil} \sum_{m=1}^{\lceil \frac{M}{2} \rceil} A_i[m] \quad \text{where} \quad A_i[m] > \text{median}(A_i),
\end{split}
\end{equation}
where $A_i$ is the list of anomaly scores $a_{i,m}~|~m=1,2,\ldots,M$ for each chunk, and where $a_i$ is the anomaly score.
This approach presumes that both stationary and non-stationary anomalies are present in at least half of the sample’s length.
For samples whose length significantly deviates from the dataset norm, we considered using quartiles to achieve a more nuanced threshold, especially when the length of the input data sample varies widely from that of the typical dataset sample.

\section{Experimental Evaluations}
\label{section:exp}
\subsection{Dataset}
We evaluated the performance of the proposed method using the \\DCASE2020 Task2 dataset, a representative ASD dataset which has also been used to evaluate other recently proposed ASD methods. 
Similar datasets include DCASE2021 Task2, DCASE2022 Task2, and DCASE2023 Task2~\cite{Kawaguchi2021,Dohi2022,Dohi2023}, however these subsequent datasets focus on domain shift control, therefore the DCASE2021 Task2 and later datasets have a hierarchical structure which is not necessarily segmented to reflect the audio data of particular target machines. 
This difference complicates the discussion when evaluating the proposed method and examining the factors that led to performance improvement.
Therefore, we used DCASE2020 Task2 data in this study, since its focus is improving ASD performance through the use of anomalous data.

The DCASE2020 Task2 dataset consists of two datasets, MIMII~\cite{Purohit_DCASE2019_01} and ToyADMOS~\cite{Koizumi_WASPAA2019_01}, both of which consist of machine operation and environmental sounds. 
The machine operation sounds were collected in a quiet environment under several different sound collection conditions, while the environmental sounds were collected in a real factory under several different conditions. 
The task organizers mixed these two datasets at various SNRs to generate audio samples. Each sample was 10 seconds long and was collected on one channel at 16,000 Hz.
Six machine types were used: Fan, Pump, Slider, and Valve from MIMII, and ToyCar and ToyConveyor from ToyADMOS. 
We evaluated our proposed method using the DCASE2020 Task2 development set, which is the same evaluation setting used in most previous studies~\cite{stgrammfn2022,Wilkinghoff2021a,Chen2023wavenet,geco2023,Guan2023,choi2024noisy}.
DCASE2020 Task2 evaluation set was used for development purposes, to determine the hyperparameters of the model.
The data for each machine ID contains about 1,000 samples of normal sounds as training data, and 200 samples each of normal and anomalous sounds as test data. 
There are three or four varieties of anomalous sounds for each machine type, but it is impossible to determine from the evaluation set what type of anomaly is contained in the anomalous data.
When training each of the 41 anomaly detectors $h$, feature vectors obtained from all of the normal data for each machine ID were utilized as training data for the respective detectors.

\begin{table*}[h]
\centering
\caption{Average performance of compared methods when using the DCASE2020 Task2 dataset. Values represent $\mathrm{aAUC}$ (the mean of AUC and partial AUC $(p=0.1)$).}
\label{table:result_aauc}
\resizebox{\textwidth}{!}{
\begin{tabular}{l|llllll|l}
\hline
Methods                     & \multicolumn{1}{c}{Fan}     & \multicolumn{1}{c}{Pump}    & \multicolumn{1}{c}{Slider}  & \multicolumn{1}{c}{Valve}   & \multicolumn{1}{c}{ToyCar}  & \multicolumn{1}{c|}{ToyConveyor} & \multicolumn{1}{c}{Average} \\ \hline
GMADE+SSC~\cite{Giri2020b,Giri2020a} & 80.65 & 83.27 & 93.41 & 91.21 & 92.72 & 73.29 & 86.27 \\
DDCSAD+BCE~\cite{kuroyanagi2021anomalous}                  & 83.42$\pm$0.26              & 83.01$\pm$0.57              & 87.27$\pm$0.97              & 97.90$\pm$0.83              & 87.72$\pm$0.42              & 59.72$\pm$0.97                   & 83.17$\pm$1.58              \\
STgram-MFN~\cite{stgrammfn2022}                  & 91.51                       & 86.85                       & 98.58                       & 99.04                       & 91.06                       & 69.09                            & 89.35                       \\
SCAdaCos~\cite{Wilkinghoff2021a} & 86.63$\pm$0.34 & 90.96$\pm$0.46 & 99.08$\pm$0.17 & 93.33$\pm$1.08 & 95.44$\pm$0.06 & 70.93$\pm$0.32      & 89.39$\pm$0.23 \\

SW-WaveNet~\cite{Chen2023wavenet}                  & 94.54                       & 84.98                       & 96.77                       & 98.14                       & 92.85                       & 74.70                            & 90.33                       \\
CLP-SCF~\cite{Guan2023}                      & 95.11              & 91.18                       & 98.65                       & 99.70                       & 93.02                       & 69.00                            & 91.12                       \\
Noisy-ArcMix~\cite{choi2024noisy}                        & \textbf{96.83}                       & 90.72                       & 98.52                       & \textbf{99.85}                       & 93.44                       & 72.53                            & 91.98                       \\
Serial-OE (Proposed)                   & 93.29$\pm$0.07              & \textbf{94.87$\pm$0.13}     & \textbf{99.56$\pm$0.17}     & 99.19$\pm$0.10     & \textbf{96.54$\pm$0.12}     & \textbf{77.77$\pm$0.49}          & \textbf{93.54$\pm$1.00}     \\ \hline
\end{tabular}
}
\end{table*}

\begin{table*}[h]
\centering
\caption{Stability of performance compared methods when using DCASE2020 Task2 dataset. 
Values represent $\mathrm{mAUC}$ (AUC of the lowest-performing machine ID of each machine type).}
\label{table:result_mauc}
\resizebox{\textwidth}{!}{
\begin{tabular}{l|llllll|l}
\hline
\multicolumn{1}{c|}{Methods} &
  \multicolumn{1}{c}{Fan} &
  \multicolumn{1}{c}{Pump} &
  \multicolumn{1}{c}{Slider} &
  \multicolumn{1}{c}{Valve} &
  \multicolumn{1}{c}{ToyCar} &
  \multicolumn{1}{c|}{ToyConveyor} &
  \multicolumn{1}{c}{Average} \\ \hline
\multicolumn{1}{c|}{DDCSAD+BCE~\cite{kuroyanagi2021anomalous}} & 69.58$\pm$1.95 & 71.33$\pm$0.93 & 79.12$\pm$1.89 & 95.19$\pm$3.02 & 74.90$\pm$1.13 & 54.96$\pm$1.27 & 74.18$\pm$1.71 \\
\multicolumn{1}{c|}{STgram-MFN~\cite{stgrammfn2022}} & 81.39          & 83.48          & 98.22          & 98.83          & 83.07          & 64.16          & 84.86          \\
SCAdaCos~\cite{Wilkinghoff2021a}    & 76.55$\pm$0.93 & 87.83$\pm$0.38 & 99.09$\pm$0.18 & 90.31$\pm$1.16 & 94.24$\pm$0.26 & 69.25$\pm$0.71 & 86.88$\pm$0.32 \\
CLP-SCF~\cite{Guan2023}                         & 88.27 & 87.27          & 98.28          & 99.58 & 86.87          & 65.46          & 87.62          \\
Noisy-ArcMix~\cite{choi2024noisy}    & \textbf{92.67} & 91.17          & 97.96          & \textbf{99.89} & 88.81          & 68.18          & 89.78          \\
Serial-OE  (Proposed) &
  84.12$\pm$0.33 &
  \textbf{94.17$\pm$0.24} &
  \textbf{99.54$\pm$0.17} &
  99.13$\pm$0.19 &
  \textbf{94.81$\pm$0.42} &
  \textbf{70.48$\pm$1.41} &
  \textbf{90.38$\pm$1.39} \\ \hline
\end{tabular}
}
\end{table*}

\subsection{Experimental Conditions}
For the sake of specific discussion, this subsection describes the development of a model of a machine type "Fan."
Note that we have implemented the same model for other machine types as well and used the same hyperparameters for all of them to simplify comparison.
When developing an ASD model for a Fan, the normal data includes all of the normal data for all seven of the Fan machine IDs. 
The pseudo-anomalous data consists of the normal sounds of all of the Pump, Slider, Valve, ToyCar, and ToyConveyor machines. 

First, as a pre-processing step, the mean and variance for the amplitude of all of the normal Fan data were calculated, and the amplitude of the entire training dataset was normalized.
The normalized data was then randomly split into chunks of $S = 2.0$ seconds, and features were extracted using a Mel filter bank with 224 frequency bins, a window size of 128 ms, and a hop size of 16 ms, covering a frequency range from 50--7,800 Hz
We analyzed the spectrograms of the training data and found that the machine operating sounds were concentrated in the higher frequency range. We applied a band-pass filter to ensure that our model focuses on the most relevant features, the frequency range where the power was concentrated (50--7,800 Hz) across all machines.
The obtained features were applied to EfficientNet-b0~\cite{xie2020self}, a powerful convolutional neural network (CNN)-based feature extractor pretrained using ImageNet~\cite{imagenet}, to obtain a $D = 128$ dimension feature vector.
When training EfficientNet-b0, the learning rate was set to 0.001, the optimization method was AdamW~\cite{adamw2019}, the scheduler was OneCycleLR~\cite{onecyclelr2019}, the number of epochs was 100, the size of a mini-batch was 128, $\alpha$ in Eq.~\ref{eq:l} was set at 10.0, and the hyperparameters of Mixup’s beta distribution $\mathcal{B}(\beta, \beta )$ were set to $\beta = 0.2$.
After training the feature extractor, the GMM of anomaly detector $h$ was trained.
Experiments using the DCASE2020 Task2 evaluation set suggested setting the number of GMM mixtures to 2. 
During inference, an $L = 10$ second sound source was divided into ten segments of $S = 2.0$ seconds, allowing for overlap.
The hyperparameters were used, which showed the highest performance in the DCASE2020 Task2 evaluation set. 
Finally, an aggregator was applied to all of the segments to compute an anomaly score for the target sound source.
\subsection{Evaluation of ASD Performance}
The performance of the proposed method was compared with the performance of methods similar to the proposed method, as well as with the performance of recently proposed state-of-the-art methods. The methods used in our performance comparison can be briefly described as follows:
\begin{list}{}{}
    \item[{\bf GMADE+SSC}~\cite{Giri2020b,Giri2020a}]  is the method that came first in the \\
    DCASE2020 Task2 Challenge~\cite{Koizumi_DCASE2020_01}. The anomaly scores for this method are calculated by ensemble, using the anomaly scores obtained by the method that uses a masked autoencoder for density estimation for each frame (group) of the mel spectrogram (GMADE)~\cite{Giri2020b} and the anomaly scores obtained by the method that performs two self-supervised classifications for machine ID classification and data augmentation (SSC)~\cite{Giri2020a}.
    \item[{\bf DDCSAD+BCE}~\cite{kuroyanagi2021anomalous}] uses the normal data of the target machine's target machine ID as normal data. This method uses normal data other than the target machine ID of the target machine and normal data other than the target machine as pseudo-abnormal data.
    It is important to note that the definitions of ‘normal data’ and ‘pseudo-anomalous data’ differ from those used in the proposed method.
    \item[{\bf STgram-MFN}~\cite{stgrammfn2022}] utilizes features obtained by classifying machine IDs using spectrograms and T-grams extracted using a 1D CNN.
    \item[{\bf SCAdaCos}~\cite{Wilkinghoff2021a}] is a method which trains the feature extractor by classifying the data according to the machine IDs of various machines. The feature vectors generated through this process are then utilized in a GMM. The training of the feature extractor leverages a metric known as Sub-Cluster AdaCos, which is combined with Mixup to produce a feature space optimally suited for ASD tasks. This approach uses a single model to extract feature spaces for each of the machines. A GMM is trained individually for each machine ID to calculate anomaly scores, allowing for tailored anomaly detection for individual machines. The feature extractor’s training involves using 16 subclusters, a number that, along with an equivalent number of GMM mixtures, yielded the best performance when using the development data, so we adopted it. All other training conditions are the same as those set in the original study~\cite{Wilkinghoff2021a}.
    \item[{\bf SW-WaveNet}~\cite{Chen2023wavenet}] is a method that integrates spectrograms and wavegrams, the latter being derived from WaveNet, to extract additional features using WaveNet itself. This approach is unique since it is the first to employ WaveNet not as a conventional generative model but as a feature extractor, as WaveNet is able to extract more detailed features from audio data, improving the model’s sound analysis ability.
    \item[{\bf  CLP-SCF}~\cite{Guan2023}] adopts a two-phase training methodology. In the initial phase the model is pretrained using a contrastive framework, which involves the application of contrastive learning to utilize the relationships between the sounds produced by a particular machine type and the sounds of the individual machine IDs within that machine type. In the next phase, the model is fine-tuned using CLP-SCF, which introduces self-supervised classification in order to enhance the process of learning the sound features pertinent to ASD. STgram-MFN is utilized as the feature extractor, enabling the model to identify the characteristics of sound data.
    \item[{\bf Noisy-ArcMix }~\cite{choi2024noisy}] improves the compactness of the within-class distribution by classifying samples virtually synthesized by Mixup to be close to the normal data distribution by ArcFace~\cite{ArcFace}. In addition, a new input feature, the temporal attention log-Mel spectrogram (TAgram), derived from the temporal attention block, is introduced into STgram-MFN.
\end{list}

To compare the average ASD performance of each method, we used average AUC (aAUC), which is the mean of the area under the receiver operating characteristic curve (AUC) and partial AUC ($p = 0.1$).
To compare stability of performance across machine IDs, which has also been examined in previous studies~\cite{stgrammfn2022,Guan2023,Dohi2022}, we used minimum AUC (mAUC) which represents the lowest AUC of a machine ID within a particular machine type. 
Tables~\ref{table:result_aauc} and~\ref{table:result_mauc} compare the performance of each method.
Results for the proposed Serial-OE method were obtained by the authors when using the DCASE2020 Task2 development dataset, while the performances of GMADE+SSC~\cite{Giri2020b,Giri2020a}, STgram-MFN~\cite{stgrammfn2022}, SW-WaveNet~\cite{Chen2023wavenet}, CLP-SCF~\cite{Guan2023}, and Noisy-ArcMix~\cite{choi2024noisy} are the results as reported in the respective studies.
The performances of DDCSAD+BCE~\cite{kuroyanagi2021anomalous}, SCAdaCos~\cite{Wilkinghoff2021a}, and the proposed method were calculated five times using different seeds, and the mean and standard deviations of the results are shown.

Table~\ref{table:result_aauc} shows that the proposed method outperformed\\ GMADE+SSC for all of the machine types. 
The single system performed better than the ensemble method that combined multiple systems.
Table~\ref{table:result_aauc} also shows that the proposed method outperformed all of the conventional methods for all of the machine types except Valve and Fan.
The operation sound of the Valve is a clicking sound, and the operation sound of the Fan is the sound of a propeller cutting through the wind, so there are almost no similarities between the two machines. Compared to the best performance of the conventional method, Noisy-ArcMix, Valve's performance difference is 0.66 points in aAUC, which is slight, and aAUC is also relatively high at 99.19\%. 
While Fan's performance difference is 3.54 points in aAUC, indicating a moderate difference, with an aAUC of 93.29\%.
The sound of the Fan is the sound of a propeller cutting through the wind, and compared to other machines, it is difficult to distinguish between the operation sound and environmental noise.
Investigating the adverse effects of Fan on the proposed method is a future issue. 
Although compared to Noisy-ArcMix, the proposed method improves aAUC by 1.7\% on average.
When comparing the proposed method's performance with that of SCAdaCos, which is similar to the proposed method, the proposed method outperformed SCAdaCos in terms of aAUC for all machine types.
Noisy-ArcMix, SCAdaCos, and the proposed method are the same in that they are trained to detect trivial changes in normal data using Mixup for the classification task. 
The proposed method differs because it trains one model for one machine type and devises a sampling strategy where irrelevant machine types are regarded as pseudo-anomalous data. 
This sampling strategy is considered effective when compared to the strategy of training one model for all machine types in that it not only ignores background noise specific to the machine being monitored but also explicitly identifies mixed sounds as anomalies by being specific to the machine being monitored.

Table~\ref{table:result_mauc} shows the stability of the performance of each method across machine IDs. 
Even if the reported aAUC is high, if the mAUC is low it indicates that some anomalous data could not be detected for a specific machine ID. 
Therefore, if a method’s mAUC is high, its ASD performance can be evaluated as stable. 
Compared to Noisy-ArcMix, which was also the best performing conventional method in this stability evaluation, the proposed method achieved an average mAUC that was 0.7\% higher.
Compared to SCAdaCos, a similar method to the proposed method, the proposed method also achieved higher stability performance in mAUC for all machine types.

These results show that the proposed method performs well not only on average, but also across many machine IDs.
All the methods shown in Tables~\ref{table:result_aauc} and~\ref{table:result_mauc}, except for DDCSAD+BCE and Serial-OE, are based on the assumption that only normal data is used. 
Therefore, even if anomalous data can be obtained, it is necessary to incorporate additional techniques to use them. 
The proposed method is superior because it can easily and efficiently handle anomalous data. 
This point is discussed in the subsection~\ref{subsection:use_anomaly}.

\begin{table*}[!t]
\centering
\caption{Average performance of the proposed method when using various loss functions for $\mathcal{L}_\mathrm{id}$ for DCASE2020 Task2 dataset. 
Values represent $\mathrm{aAUC}$ (the mean of AUC and partial AUC $(p=0.1)$).}
\label{table:result_l_id}
\resizebox{\textwidth}{!}{
\begin{tabular}{l|llllll|l}
\hline
Loss function &
  \multicolumn{1}{c}{Fan} &
  \multicolumn{1}{c}{Pump} &
  \multicolumn{1}{c}{Slider} &
  \multicolumn{1}{c}{Valve} &
  \multicolumn{1}{c}{ToyCar} &
  \multicolumn{1}{c|}{ToyConveyor} &
  \multicolumn{1}{c}{Average} \\ \hline
BCE (Proposed) &
  \textbf{93.29$\pm$0.07} &
  \textbf{94.87$\pm$0.13} &
  \textbf{99.56$\pm$0.17} &
  \textbf{99.19$\pm$0.10} &
  \textbf{96.54$\pm$0.12} &
  \textbf{77.77$\pm$0.49} &
  \textbf{93.54$\pm$1.00} \\
Cross-entropy & 88.72$\pm$0.21 & 91.42$\pm$0.38 & 98.60$\pm$0.25 & 96.37$\pm$0.66 & 92.12$\pm$0.79 & 69.29$\pm$0.32 & 89.42$\pm$0.18 \\
SCAdaCos~\cite{Wilkinghoff2021a}     & 84.83$\pm$0.52 & 88.72$\pm$0.59 & 97.81$\pm$0.68 & 88.97$\pm$2.68 & 84.11$\pm$0.79 & 63.38$\pm$1.01 & 84.64$\pm$0.46 \\
ArcFace~\cite{ArcFace} & 85.09$\pm$0.19 & 85.00$\pm$0.83 & 91.18$\pm$4.77 & 91.43$\pm$2.58 & 66.33$\pm$2.12 & 65.50$\pm$0.60 & 80.75$\pm$1.04  \\ \hline

\end{tabular}
}
\end{table*}
\begin{figure}[!t]
\centering
\subfloat[Binary cross-entropy]{
\begin{minipage}[t]{0.47\columnwidth}
\centering
\includegraphics[clip, width=\linewidth]{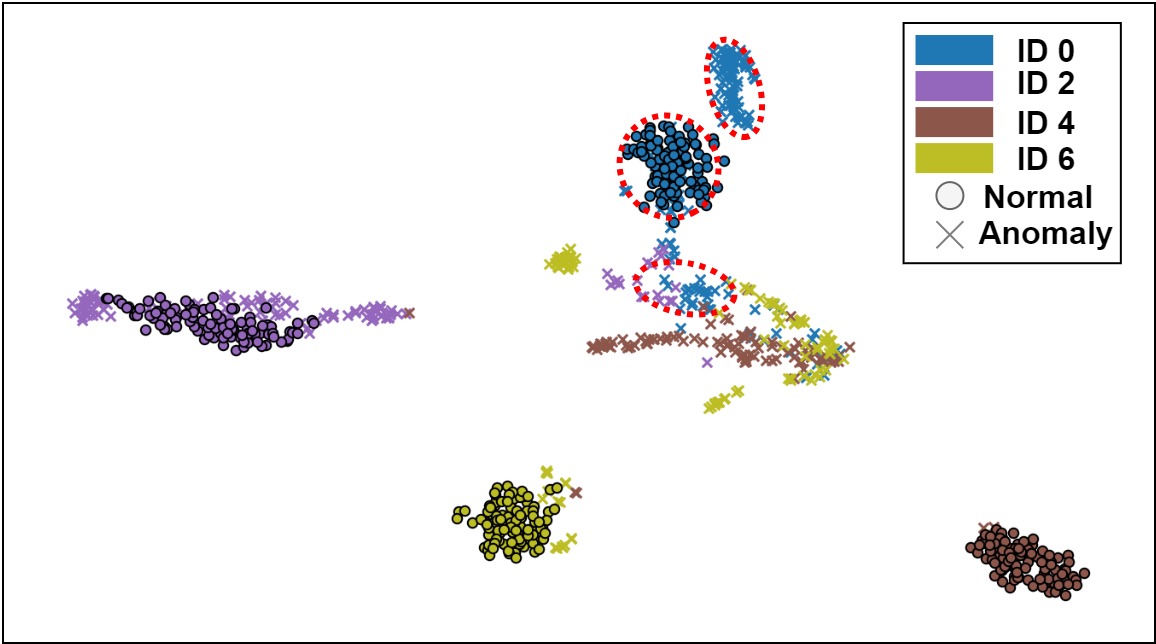}
\label{fig:l_bce}
\end{minipage}
}
\hfill
\subfloat[Cross-entropy]{
\begin{minipage}[t]{0.47\columnwidth}
\centering
\includegraphics[clip, width=\linewidth]{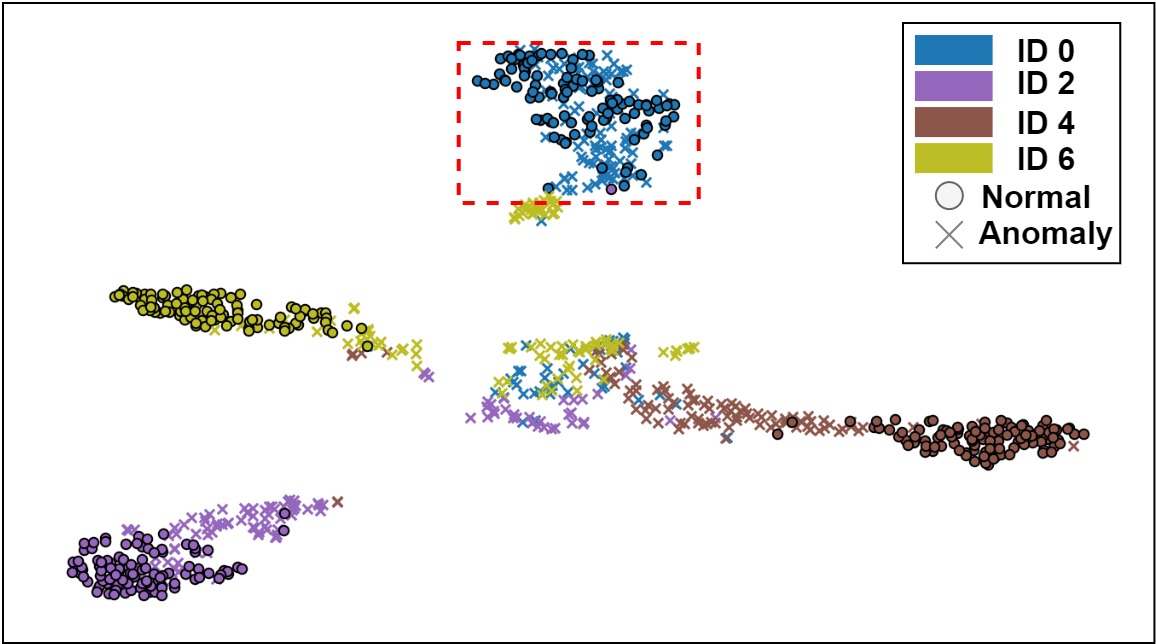}
\label{fig:l_ce}
\end{minipage}
}
\\[1em]
\subfloat[SCAdaCos]{
\begin{minipage}[t]{0.47\columnwidth}
\centering
\includegraphics[clip, width=\linewidth]{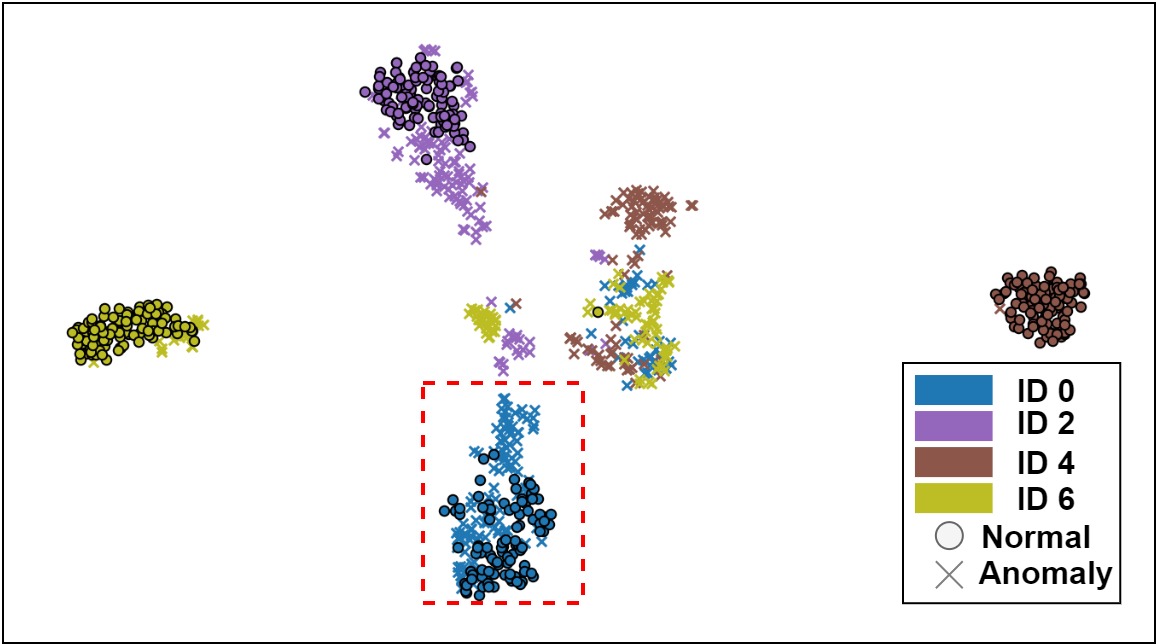}
\label{fig:l_scadacos}
\end{minipage}
}
\hfill
\subfloat[ArcFace]{
\begin{minipage}[t]{0.47\columnwidth}
\centering
\includegraphics[clip, width=\linewidth]{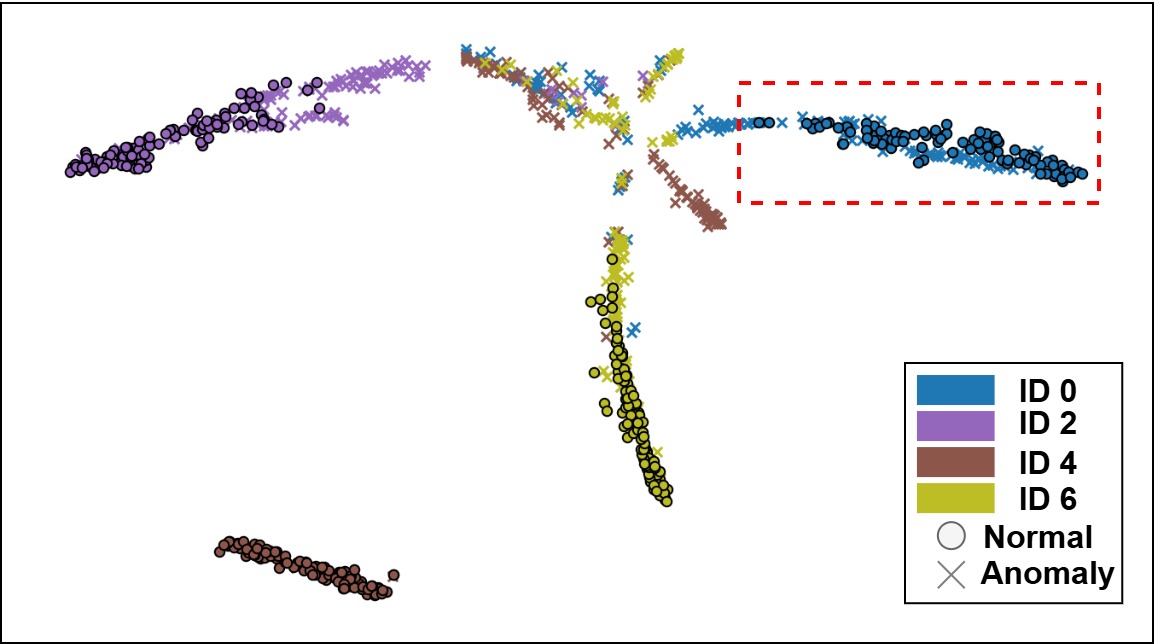}
\label{fig:l_arcface}
\end{minipage}
}
\caption{
Visualizations of data for machine type Fan using t-SNE, when varying loss function $\mathcal{L}_\mathrm{id}$ of the proposed method. (a) Feature space distribution obtained using BCE. (b) Feature space distribution obtained using cross-entropy. (c) Feature space distribution obtained using SCAdaCos~\cite{Wilkinghoff2021a}. (d) Feature space distribution obtained using ArcFace~\cite{ArcFace}. The symbols $\circ$ and $\times$ denote normal and anomalous sounds, respectively. The normal and anomalous feature space distributions for machine ID 0 are encircled by red dotted lines.
}
\label{fig:tsen}
\end{figure}

\subsection{Ablation studies}
To investigate the reasons for the superior performance achieved by the proposed method, we performed ablation studies using aAUC as the evaluation metric in order to compare the average performance of each ASD method. To allow qualitative evaluation, the feature space was visualized using t-SNE~\cite{tsen}.

\subsubsection{Evaluation of loss functions used for machine ID classification}
\label{subsubsection:l_id}
Table~\ref{table:result_l_id} shows the results in terms of aAUC when the feature extractor was trained using each of the loss functions. 
Fig.~\ref{fig:tsen} shows t-SNE visualizations (perplexity$=40$) of the feature space obtained when using each loss function with the Fan DCASE2020 Task2 development set. 
Table~\ref{table:result_l_id} and Fig.~\ref{fig:tsen} compare the most suitable loss function for $\mathcal{L}_\mathrm{id}$ in Eq.~\ref{eq:l_id}.
GMADE+SSC was excluded from the analysis because it is an ensemble system.
The proposed method and DDCSAD+BCE use BCE as the loss function for machine ID classification, and SCAdaCos uses SCAdaCos as the loss function for machine ID classification. 
The other methods, STgram-MFN, SW-WaveNet, CLP-SCF, and Noisy-ArcMix, use ArcFace as the loss function for machine ID classification.
We compared the proposed method using these loss functions for machine ID classification previously employed in conventional methods as Eq.~\ref{eq:l_id}.
We experimented with the proposed method by changing the BCE used for $\mathcal{L}_\mathrm{id}$ in Eq.~\ref{eq:l_id} to cross-entropy, SCAdaCos, and ArcFace to evaluate the differences in the resulting feature spaces.
Note that Mixup and $\mathcal{L}_\mathrm{type}$ in Eq.~\ref{eq:l_type} were also applied in all versions of the system.
The samples affected by changing the loss function in Eq.~\ref{eq:l_id} contain two types of sounds: mixtures of normal data and normal data, and mixtures of normal data and pseudo-anomalous data.
Focusing on these samples, we discuss the effects of using each loss function.
BCE acquires features by focusing on whether the input sound contains the sound of the target machine ID or not, cross-entropy, SCAdaCos, and ArcFace acquire features for classification based on the class to which the input sample belongs, and ArcFace and SCAdaCos acquire features that minimize intra-class variance, and SCAdaCos uses sub-clusters, so that a machine ID which has several different features can be classified according to each feature. 
Cross-entropy, SCAdaCos, and ArcFace require the sum of the probabilities of each class to be one, due to the characteristics of these loss functions.
Therefore, when normal and pseudo-anomalous data were mixed, the probability of the normal data was set to 1.
The usual mixing ratio of each sample was used when mixing samples containing normal data and normal data.

The results provided in Table~\ref{table:result_l_id} show that the best ASD performance was obtained for all machine types when BCE was used as the loss function for $\mathcal{L}_\mathrm{id}$ in the proposed method.
We will first discuss the reasons for the higher performance of BCE compared to cross-entropy. 
Comparing the visualization when using BCE in Fig.~\ref{fig:l_bce} and the visualization when using cross-entropy in Fig.~\ref{fig:l_ce}, we can see that the classes of normal data are more tightly clustered in the BCE visualization at ID 0, shown circled with a red dotted line. 
Considering the characteristics of the loss function, it is thought that using features obtained by focusing on whether or not a particular machine ID is included is more effective for ASD than using features obtained by classifying machine IDs, as in the case of cross-entropy.

Next, we compare the performance of the proposed method when using SCAdaCos and ArcFace as the loss function for $\mathcal{L}_\mathrm{id}$.
Table~\ref{table:result_l_id} shows that ASD performance decreases when either SCAdaCos or ArcFace is used.
Comparing the visualizations for BCE in Fig.~\ref{fig:l_bce}, and for SCAdaCos in Fig.~\ref{fig:l_scadacos} and ArcFace in  Fig.~\ref{fig:l_arcface}, we see that both SCAdaCos and ArcFace visualizations contain more anomalous data in the normal data cluster for machine ID 0 (red dotted line) than when using BCE.
Since both SCAdaCos and ArcFace utilize $\mathcal{L}_\mathrm{type}$ to perform classification, using the norm of the feature vectors, its lower performance may be due to the increased difficulty of performing angle-based classification.
Therefore, we need to carefully combine a loss function using the norm of feature vectors with angle-based metric training~\cite{Wilkinghoff2024}.

These results suggest that which loss function is used when training the feature extractor is important for distinguishing whether the sound of the target machine is included in the input sound.

\subsubsection{Evaluation of training methods for feature extractors}
\begin{table*}[]
\centering
\caption{Average performance of ablation studies for elements other than $\mathcal{L}_\mathrm{id}$ of the proposed method when using the DCASE2020 Task2 dataset.
Values represent $\mathrm{aAUC}$ (the mean of AUC and partial AUC $(p=0.1)$).
}
\label{table:result_element}
\resizebox{\textwidth}{!}{
\begin{tabular}{l|>{\centering\arraybackslash}p{1.2em}>{\centering\arraybackslash}p{2.7em}>{\centering\arraybackslash}p{2.9em}|>{\centering\arraybackslash}p{4.8em}>{\centering\arraybackslash}p{4.8em} >{\centering\arraybackslash}p{4.8em} >{\centering\arraybackslash}p{4.8em} >{\centering\arraybackslash}p{4.8em} >{\centering\arraybackslash}p{3em}|l}
\hline
Method &
  $L_\mathrm{type}$ &
  weight &
  Mixup &
  \multicolumn{1}{c}{Fan} &
  \multicolumn{1}{c}{Pump} &
  \multicolumn{1}{c}{Slider} &
  \multicolumn{1}{c}{Valve} &
  \multicolumn{1}{c}{ToyCar} &
  \multicolumn{1}{c|}{ToyConveyor} &
  \multicolumn{1}{c}{Average} \\ \hline
Serial-OE &
  $\checkmark$ &
  $\checkmark$ &
  \multicolumn{1}{c|}{$\checkmark$} &
  \textbf{93.29$\pm$0.07} &
  \textbf{94.87$\pm$0.13} &
  \textbf{99.56$\pm$0.17} &
  \textbf{99.19$\pm$0.10} &
  \textbf{96.54$\pm$0.12} &
  \textbf{77.77$\pm$0.49} &
  \textbf{93.54$\pm$1.00} \\
w/o $L_\mathrm{type}$ &
  \multicolumn{1}{l}{} &
  $\checkmark$ &
  \multicolumn{1}{c|}{$\checkmark$} &
  90.46$\pm$0.27 &
  92.69$\pm$0.36 &
  98.66$\pm$0.29 &
  98.01$\pm$0.45 &
  94.59$\pm$0.32 &
  74.44$\pm$1.67 &
  91.48$\pm$0.35 \\
w/o weight &
  $\checkmark$ &
  \multicolumn{1}{l}{} &
  \multicolumn{1}{c|}{$\checkmark$} &
  87.15$\pm$0.33 &
  87.96$\pm$0.57 &
  98.70$\pm$0.19 &
  95.19$\pm$0.68 &
  91.95$\pm$0.56 &
  72.50$\pm$0.22 &
  88.91$\pm$0.09 \\
w/o Mixup &
  $\checkmark$ &
  $\checkmark$ &
  \multicolumn{1}{l|}{} &
  86.52$\pm$0.37 &
  91.98$\pm$0.76 &
  98.40$\pm$0.32 &
  98.83$\pm$0.58 &
  94.30$\pm$0.44 &
  62.40$\pm$0.74 &
  88.74$\pm$0.22 \\
w/ $L_\mathrm{type}$ &
  $\checkmark$ &
  \multicolumn{1}{l}{} &
  \multicolumn{1}{l|}{} &
  85.15$\pm$0.31 &
  85.86$\pm$0.76 &
  98.76$\pm$0.24 &
  96.41$\pm$0.43 &
  91.96$\pm$0.22 &
  58.28$\pm$0.28 &
  86.07$\pm$0.11 \\
w/ weight &
  \multicolumn{1}{l}{} &
  $\checkmark$ &
  \multicolumn{1}{l|}{} &
  83.08$\pm$0.55 &
  84.93$\pm$0.20 &
  97.08$\pm$1.01 &
  99.02$\pm$0.39 &
  92.07$\pm$0.70 &
  61.16$\pm$0.58 &
  86.22$\pm$0.35 \\
w/ Mixup &
  \multicolumn{1}{l}{} &
  \multicolumn{1}{l}{} &
  \multicolumn{1}{c|}{$\checkmark$} &
  87.05$\pm$0.66 &
  88.12$\pm$0.40 &
  97.48$\pm$0.28 &
  98.91$\pm$0.13 &
  91.48$\pm$0.63 &
  70.76$\pm$0.31 &
  88.97$\pm$0.12 \\
only $L_\mathrm{id}$ &
  \multicolumn{1}{l}{} &
  \multicolumn{1}{l}{} &
  \multicolumn{1}{l|}{} &
  82.02$\pm$0.54 &
  84.80$\pm$0.34 &
  97.77$\pm$0.61 &
  99.34$\pm$0.14 &
  92.44$\pm$0.29 &
  56.80$\pm$0.63 &
  85.53$\pm$0.11 \\ \hline
\end{tabular}
 }
\end{table*}

Table~\ref{table:result_element} shows the results of our comparison of the performance of the proposed method when each element incorporated during training of the feature extractor is deleted.
First, focusing on the average, it is suggested that all three of these elements are effective because the performance of the proposed method is degraded when any of the elements are deleted compared to the proposed method, and all methods improve performance compared to only $\mathcal{L}_\mathrm{id}$.
In addition, when comparing the performance of the two elements combined and the performance of the individual elements, all combinations improved performance to the same or greater extent.
These comparisons of the impact of each factor on ASD performance reveal that Mixup has the largest positive impact on performance, followed by pre-training of the weights using ImageNet, and then loss function $\mathcal{L}_\mathrm{type}$, in that order. 

The large improvement in ASD performance when using Mixup indicates that acquiring an intermediate representation of the feature space is effective for improving performance.
This allows the distribution of the anomalous data, which has features slightly different from normal data, into clusters farther away from the normal data.

The ASD performance was also improved for all machine types by initializing the feature extractor using weights from models pre-trained on the ImageNet classification task.
This finding aligns with previous studies that have reported improved performance using pre-trained models~\cite{Roth2022,Kyrill}.
Our results also show that the proposed method effectively utilizes weights from models trained on other classification tasks as initial values.
However, \cite{Wilkinghoff2023e} reported no performance improvement when using pre-trained models. Thus, the effect of using pre-trained weights may depend on the specific method employed.

Finally, focusing on the effect of loss function $\mathcal{L}_\mathrm{type}$, it was found that ASD performance improved when $\mathcal{L}_\mathrm{type}$  was used as a norm to classify normal and pseudo-anomalous data.
This is because $\mathcal{L}_\mathrm{type}$ concentrates anomalous data at a point near the origin, improving anomaly detection performance.
$\mathcal{L}_\mathrm{type}$ can also create intermediate representations between normal and pseudo-anomalous data when used with Mixup, which is also thought to improve performance.

The results of this ablation study indicate that Mixup, initialization with pretrained weights and $\mathcal{L}_\mathrm{type}$ used in the proposed method were all effective for improving ASD performance.

\subsubsection{Evaluation without machine ID information}
\begin{table*}[h]
\centering
\caption{Average performance of models trained without machine ID information when using the DCASE2020 Task2 dataset.
Values represent $\mathrm{aAUC}$ (the mean of AUC and partial AUC $(p=0.1)$).
}
\label{table:result_wo_id}
\resizebox{\textwidth}{!}{
\begin{tabular}{l|>{\centering\arraybackslash}p{1.2em}>{\centering\arraybackslash}p{1.4em}>{\centering\arraybackslash}p{2.6em}|>{\centering\arraybackslash}p{4.8em}>{\centering\arraybackslash}p{4.8em} >{\centering\arraybackslash}p{4.8em} >{\centering\arraybackslash}p{4.8em} >{\centering\arraybackslash}p{4.8em} >{\centering\arraybackslash}p{3em}|l}
\hline
Method &
  $\mathcal{L}_\mathrm{id}$ &
  $\mathcal{L}_\mathrm{type}$ &
  Mixup &
  \multicolumn{1}{c}{Fan} &
  \multicolumn{1}{c}{Pump} &
  \multicolumn{1}{c}{Slider} &
  \multicolumn{1}{c}{Valve} &
  \multicolumn{1}{c}{ToyCar} &
  \multicolumn{1}{c|}{ToyConveyor} &
  \multicolumn{1}{c}{Average} \\ \hline
Serial-OE &
  \textbf{$\checkmark$} &
  \textbf{$\checkmark$} &
  \textbf{$\checkmark$} &
  \textbf{93.29$\pm$0.07} &
  \textbf{94.87$\pm$0.13} &
  \textbf{99.56$\pm$0.17} &
  \textbf{99.19$\pm$0.10} &
  \textbf{96.54$\pm$0.12} &
  \textbf{77.77$\pm$0.49} &
  \textbf{93.54$\pm$1.00} \\
SCAdaCos~\cite{Wilkinghoff2021a} &
  --- &
  --- &
  $\checkmark$ &
    86.63$\pm$0.34&
    90.96$\pm$0.46&
    99.08$\pm$0.17&
    93.33$\pm$1.08&
    95.44$\pm$0.06&
    70.93$\pm$0.32&
    89.39$\pm$0.23\\
SCAdaCos (w/o ID)&
  --- &
  --- &
  $\checkmark$ &
  75.72$\pm$0.51 &
  77.10$\pm$0.61 &
  87.44$\pm$0.43 &
  76.25$\pm$0.55 &
  76.79$\pm$0.82 &
  54.83$\pm$0.27 &
  74.69$\pm$0.38 \\
w/o $\mathcal{L}_\mathrm{id}$ &
  \multicolumn{1}{l}{} &
  $\checkmark$ &
  $\checkmark$ &
  77.86$\pm$0.94 &
  81.93$\pm$1.61 &
  88.54$\pm$1.65 &
  71.77$\pm$1.21 &
  69.66$\pm$1.79 &
  54.52$\pm$0.85 &
  74.05$\pm$0.45 \\
only $\mathcal{L}_\mathrm{type}$ &
  \multicolumn{1}{l}{} &
  $\checkmark$ &
  \multicolumn{1}{l|}{} &
  75.20$\pm$1.28 &
  71.62$\pm$1.62 &
  91.81$\pm$0.43 &
  58.84$\pm$1.19 &
  61.11$\pm$1.86 &
  52.91$\pm$0.51 &
  68.58$\pm$0.50 \\ \hline
\end{tabular}
}
\end{table*}
The conventional and proposed methods are based on the assumption that training is performed using machine ID information.
This information is used because it allows the extraction of features that reflect the target machine’s normal 
data in greater detail, improving ASD performance.
Thus, it is desirable to use machine IDs or similar information when available, however such data is not always available.
Table~\ref{table:result_wo_id} shows ASD performance results when the feature extractor and anomaly detector parameters are trained without the use of machine ID information. When using the proposed method, we set $\lambda=0$ in Eq.\ref{eq:l}.
Forty-one classifications were created when SCAdaCos used machine ID information, but only six were created when only machine types were used. 
When using each method, GMMs were trained for each machine type to create six anomaly detectors. Our results in Table V show that when machine ID information is not used, ASD performance was degraded for both models for all machine types. 

As was the case with machine IDs, ASD performance improved when the feature extractor was trained to obtain an intermediate representation between normal and pseudo-anomalous data using Mixup, confirming the effectiveness of acquisition of intermediate representations.

Next, we compared the performance of the proposed method without machine IDs with that of SCAdaCos without machine IDs and found that the performance of both methods was almost the same in terms of average performance. 
When only normal data is used, and machine IDs cannot be used, there is no difference between the two methods. 
However, the proposed method has the advantage of being able to utilize anomalous data, as explained in the following subsection.
\subsection{Evaluating performance when anomalous data is utilized as pseudo-anomalous data during training}
\begin{figure}[!t]
\centering
\resizebox{0.9\columnwidth}{!}{
\begin{tabular}{cc}
\subfloat[Average AUC (aAUC)]{\includegraphics[width=0.49\columnwidth]{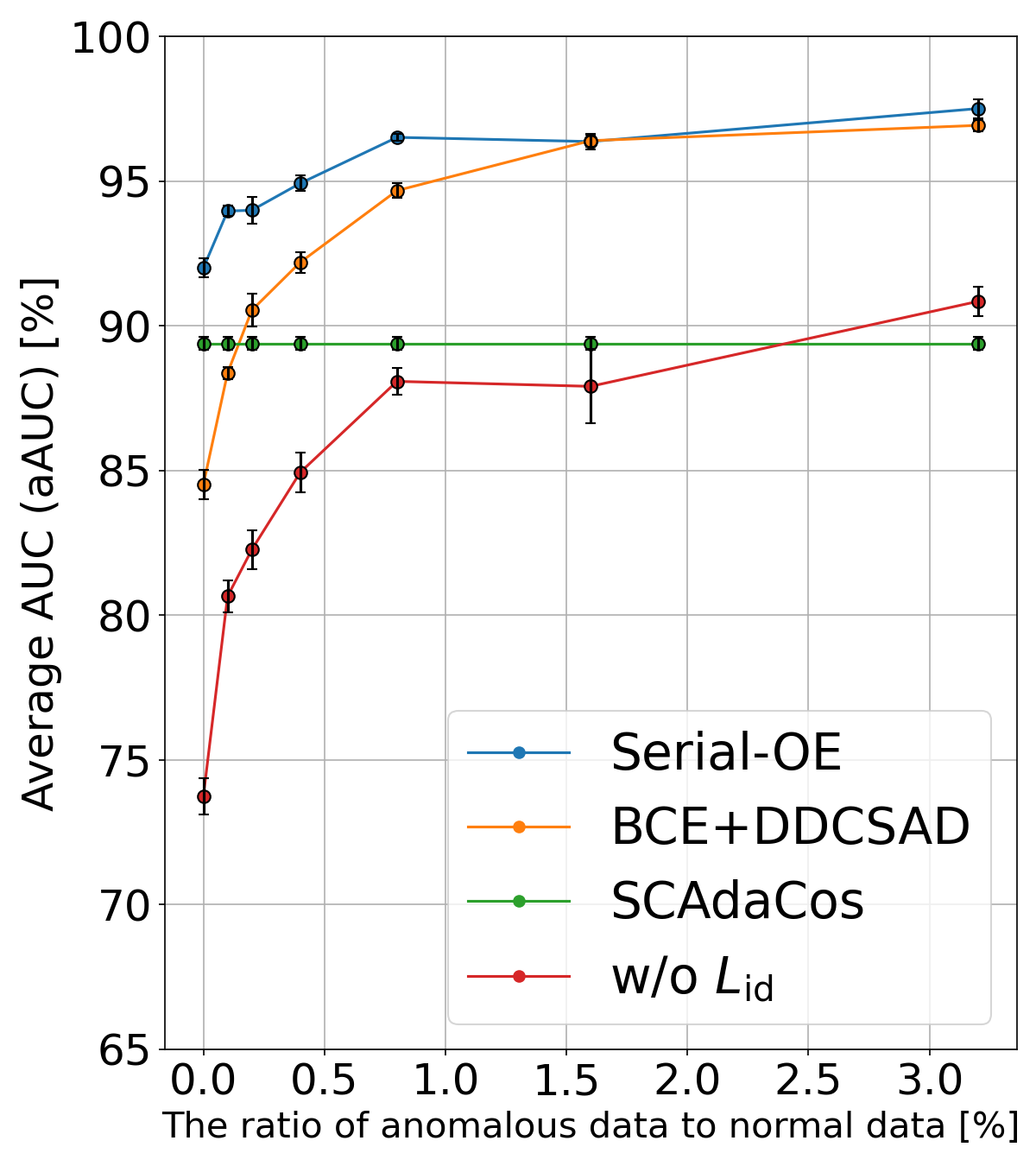}\label{fig:use_aauc}}
\hfil
\subfloat[Minimum AUC (mAUC)]{\includegraphics[width=0.49\columnwidth]{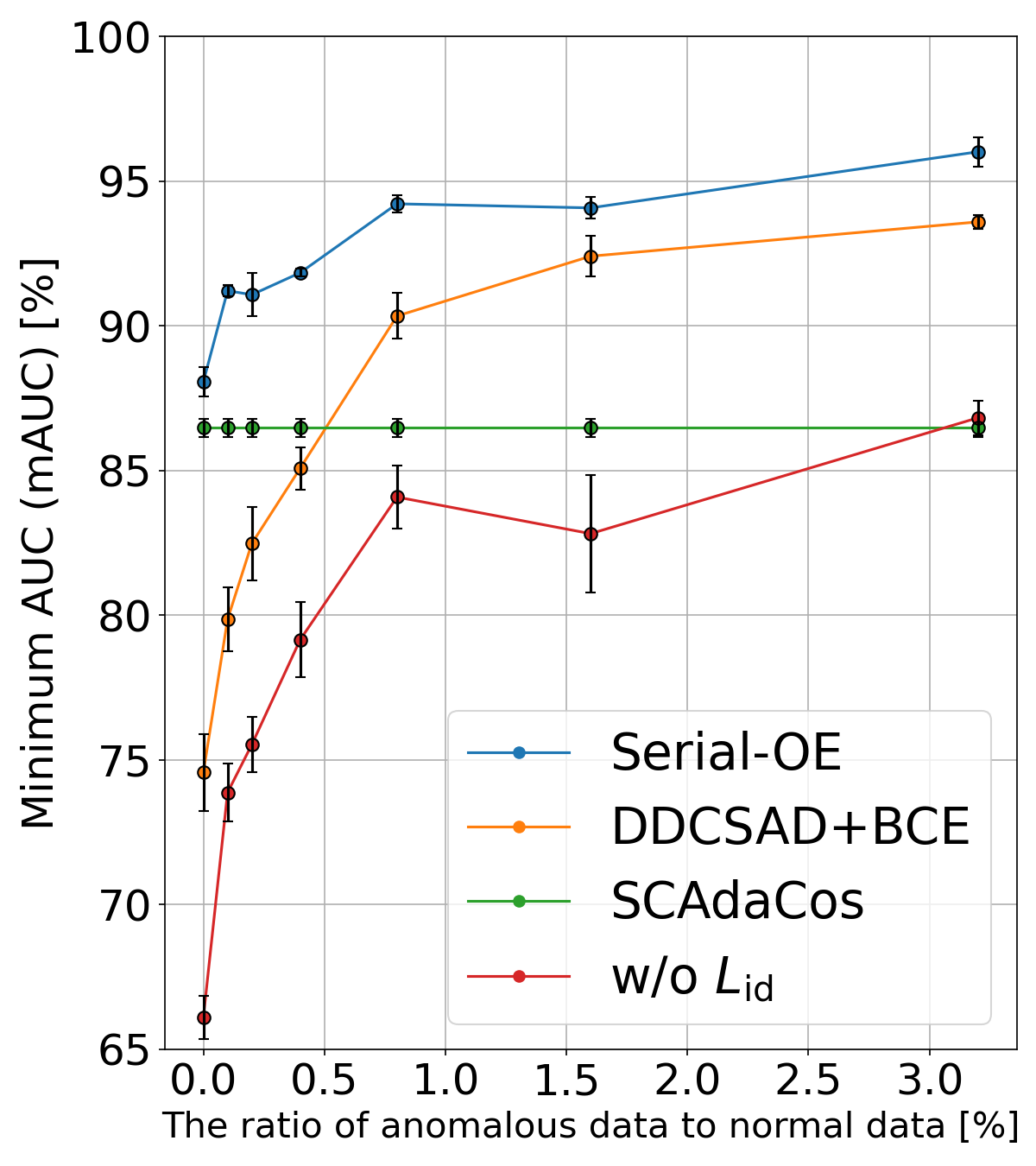}
\label{fig:use_mauc}}
\end{tabular}
}
\caption{
Relationship between the ratio of anomalous to normal data used during training and ASD performance, when using two different performance evaluation metrics: (a) aAUC [\%] and (b) mAUC [\%]. Error bars represent the standard error obtained from five calculations with different seeds.
}
\label{fig:use}
\end{figure}

Since real anomalous audio data contains features useful for detecting anomalous conditions, it is likely that it can be used to improve performance. 
Therefore, we also investigated whether the proposed method can achieve higher ASD performance by using a small amount of anomalous data during training.
We also evaluated the effectiveness of using anomalous data when information such as machine IDs is not available.

Fig.~\ref{fig:use_aauc} is a graph in which the ratio of anomalous data to training data is represented on the x-axis, while average AUC (aAUC) is represented on the vertical axis. A similar graph is shown in Fig.~\ref{fig:use_mauc}, but minimum AUC (mAUC) is shown on the vertical axis. 
Here, the volume of training data per machine ID is 1,000 samples, and each sample has a duration of 10 seconds, so a ratio of 1\% anomalous data represents  10 samples obtained from 100 seconds of anomalous data. 
In the experiment, $\{2^i~|~i=0,1,\ldots, 5\}$ samples of anomalous data were used for training.
Because the evaluation data contains approximately 100 anomalous samples, we set $i=5$ as the maximum to ensure a sufficient amount of data for evaluation.
Since the proposed method defines and binarily classifies normal and pseudo-anomalous data, as described in Section~\ref{section:proposed_method}, a small amount of real anomalous data can be used to improve ASD performance, when it is available.
The same experiment was conducted when machine ID information was unavailable. DDCSAD + BCE~\cite{kuroyanagi2021anomalous}, a conventional method for classifying normal and pseudo-anomalous data, was used for comparison.
SCAdaCos~\cite{Wilkinghoff2021a} performance scores are also shown as a reference.
Figs.~\ref{fig:use_aauc} and~\ref{fig:use_mauc} show that Serial-OE achieved the highest performance in terms of both aAUC and mAUC, respectively, followed by DDCSAD+BCE.
These results show that the proposed method can utilize anomalous data more effectively than DDCSAD+BCE to improve ASD performance, in terms of both average performance and performance stability.
The proposed method significantly improved ASD performance even when only 0.1\% of the anomalous data (1 sample of 10-second data)  was added to the normal data as training data, demonstrating the usefulness of utilizing a small amount of anomalous data during training. 
This improvement was equivalent to a 2.4\% improvement in performance in terms of aAUC, which is a much easier means of boosting performance than developing a new ASD method.

Next, we compared ASD performance when anomalous data is used for training, but the data does not contain machine IDs. 
The results in Figs.~\ref{fig:use_aauc} and~\ref{fig:use_mauc} show that when the percentage of anomalous data is 3.2\%, $i.e.$, when only 320 seconds of anomalous data is available, the proposed method’s performance is equal to or better than that of the conventional method using machine IDs (SCAdaCos). 
Many ASD methods are based on the assumption that machine ID information is a baseline characteristic of audio data used for ASD. 
However, in real environments it may not always be possible to collect data equivalent to machine IDs. 
On the other hand, anomalous data will likely be available, since machines monitored by ASD systems presumably experience anomalous states on a somewhat regular basis. Moreover, hours of anomalous data are not needed for training, since only a few seconds to a few minutes of data is sufficient to improve performance. 
Thus, the proposed method can be used to improve performance by using a small amount of anomalous data, even when machine ID information is unavailable, which is considered to be a major operational advantage.
The fact that the hyperparameters do not need to be adjusted for each machine is also considered to be an operational advantage, since this allows the model to be easily adapted if many machines are being monitored, for example.

\subsection{Evaluating performance when anomalous data used for training is contaminated with normal data}
\label{subsection:use_anomaly}
\begin{figure}[!t]
\centering
\resizebox{0.9\columnwidth}{!}{
\begin{tabular}{cc}
\subfloat[Average AUC (aAUC)]{\includegraphics[width=0.49\columnwidth]{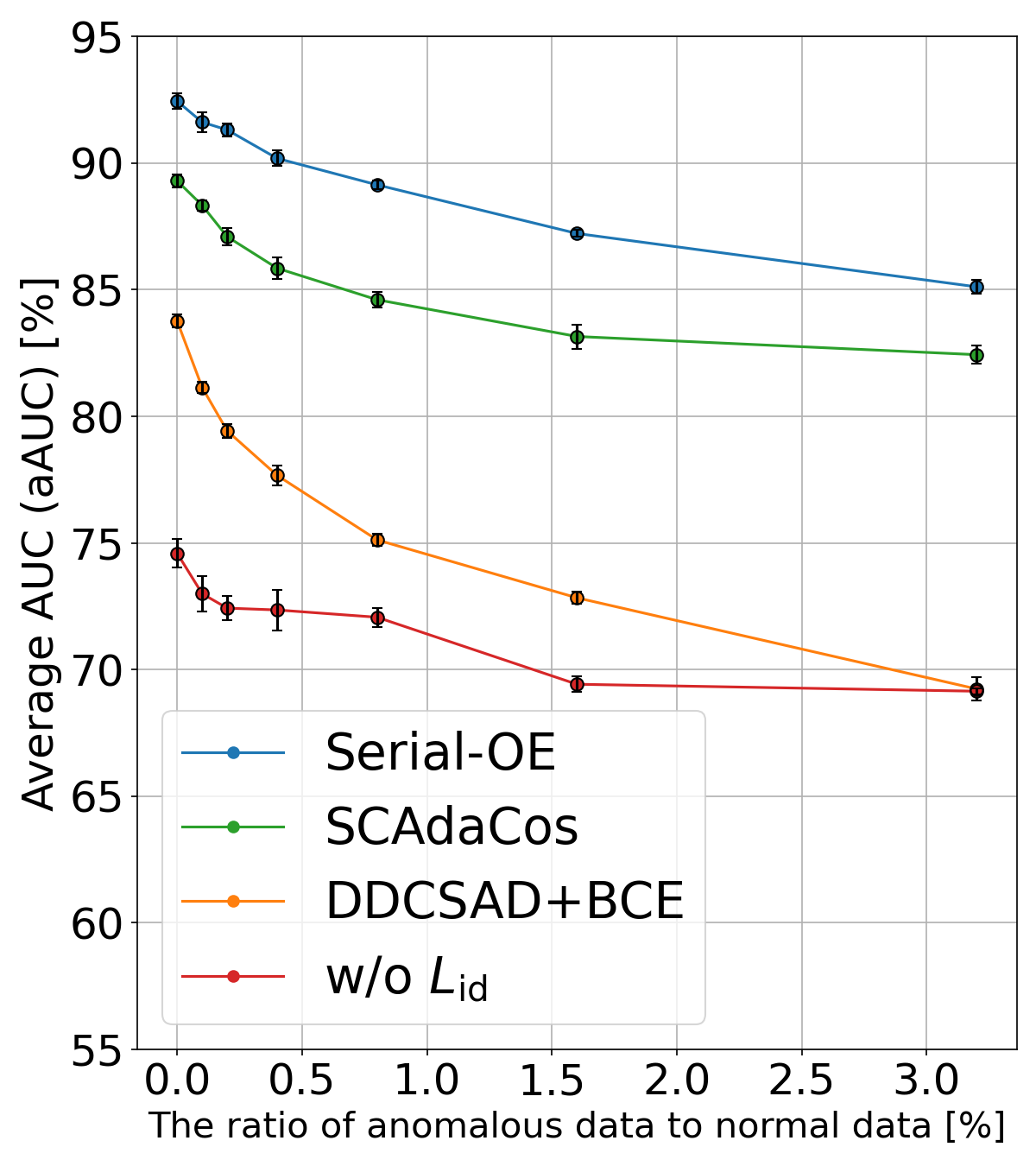}%
\label{fig:contami_aauc}}
\hfil
\subfloat[Minimum AUC (mAUC)]{\includegraphics[width=0.49\columnwidth]{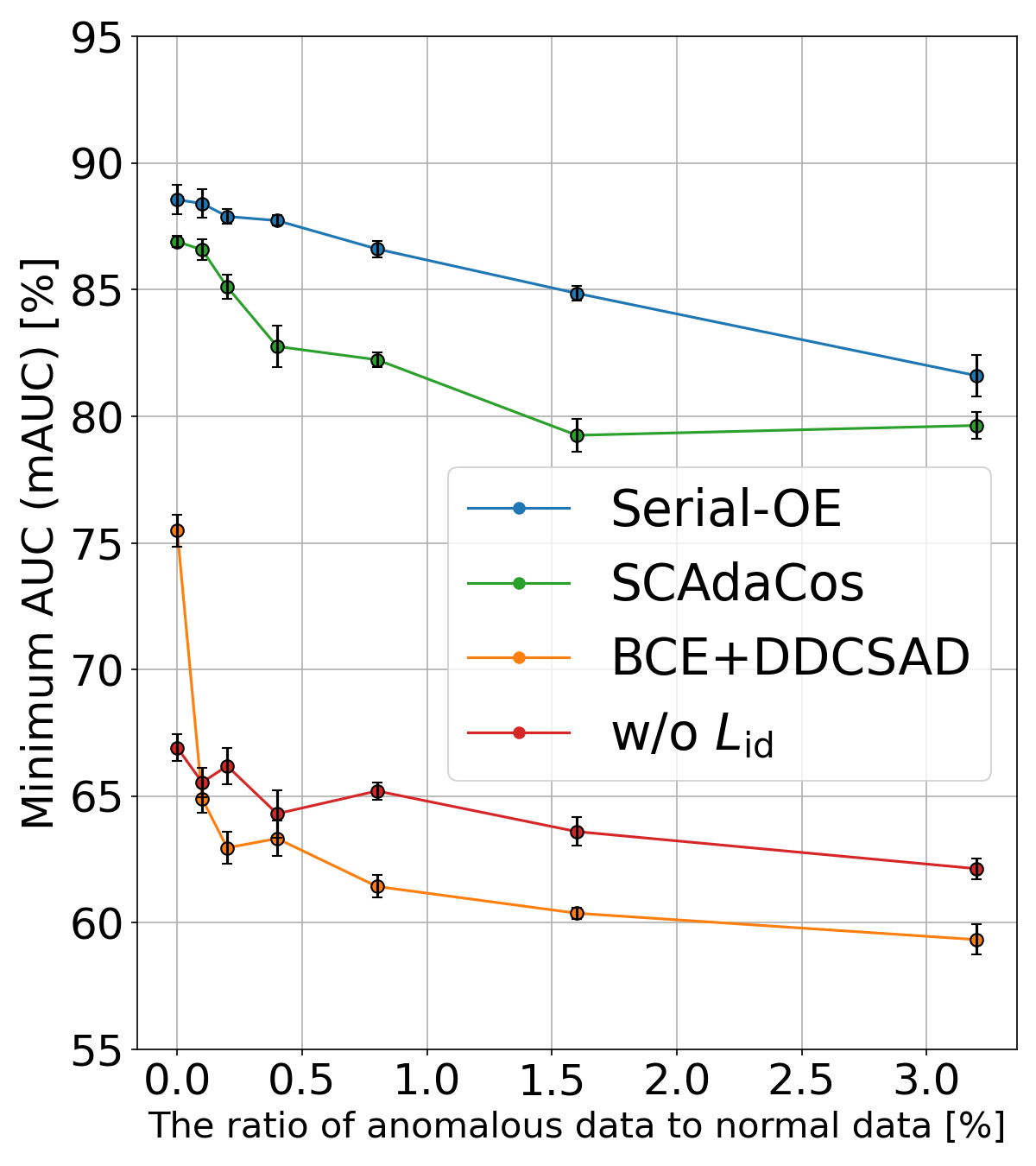}
\label{fig:contami_mauc}}
\end{tabular}
}
\caption{ Relationship between the ratio of anomalous to normal data used during training and ASD performance when the anomalous data was contaminated with various amounts of normal data, using two different evaluation metrics: (a) aAUC [\%], and (b) mAUC [\%]. Error bars represent the standard error obtained from five calculations with different seeds. }
\label{fig:contami}
\end{figure}
Finally, when utilizing a small amount of anomalous data obtained during machine operation to better train the ASD system, depending on the data collection method used, this anomalous data may contaminate the normal training data. 
Therefore, we evaluated the performance of the proposed method when the small amount of real anomalous data added to the normal data. 
We randomly contaminated the normal training data with anomalous data, then trained the feature extractor and anomaly detector using this data. 
In Fig.~\ref{fig:contami_aauc} the percentage of contaminated anomalous data used is shown on the x-axis, while performance in terms of aAUC is represented on the y-axis. 
Similarly, Fig.~\ref{fig:contami_mauc} shows the percentage of contaminated anomalous data on the x-axis and stability in terms of mAUC on the y-axis. 
Fig.~\ref{fig:contami} shows that the more contaminated the normal data was, the worse the performance of all the ASD methods. 
However, the proposed method achieved the best performance at all contamination levels compared to the other methods, thus the proposed method is more robust against contaminated anomalous data than the other methods. 
In particular, the proposed method was able to achieve the same performance as SCAdaCos, even when the normal data was contaminated with about 80 seconds of anomalous data. 
If this amount of contaminated anomalous data was successfully classified and used for training, the aAUC-based performance could be improved by nearly 5\%, as shown in Fig~\ref{fig:use_aauc}.
Therefore, the management of anomalous data is very important in terms of improving model performance.

In terms of the ratio of the decrease in anomaly detection scores in relation to the ratio of anomalous data contamination, DDCSAD+BCE had the sharpest decrease in performance. 
DDCSAD+BCE is considered to be more vulnerable to contaminated data than other methods because it trains a single model for a single machine ID.
The proposed method and SCAdaCos were less sensitive to contamination of the anomalous data than the other methods. 
Comparing the proposed method with SCAdaCos, the proposed method exhibited a more gradual decline in performance as the percentage of contamination with anomalous data increased, compared to SCAdaCos. 
The proposed method uses one model to infer one machine type, while SCAdaCos uses one model to infer all of the machine types.
The influence of contaminated normal data on each machine ID may be reduced by defining the feature space for multiple types of machines. 
This result suggests that single models used for multiple types of machines may be more robust to contamination of the anomalous data than methods which use a different model for each machine type. 
On the other hand, the method that improves performance the most when utilizing anomalous data is the one which uses a different model for each machine type, so the relationship between resistance to contamination and increased performance is considered to be a trade-off.

\section{Limitations}
\label{section:limit}
Although this study evaluated the change in ASD performance when real anomalous data was used for training, many different types of anomalous conditions exist.
Due to the limitations of the DCASE2020 Task2 dataset, this study did not investigate how effective data from various different anomalous conditions are for improving ASD performance when training the proposed method. 
Further research on this point would be helpful for finding the most suitable conditions when adopting the proposed Serial-OE method.
Furthermore, the proposed method assumes the use of machine IDs, obtained from annotated evaluation data, which is not always available. 
Since the performance of both the proposed and conventional methods decreases when machine IDs are not used, it is necessary to investigate what types of data can be used as alternative features. This issue was also not investigated due to the limitations of the dataset.

Investigating the impact of domain shift on performance, and the proposed method’s applicability to edge devices were also outside the scope of this study.
We believe that investigation of these issues would be advantageous.

\section{Conclusion}
\label{section:conclusion}
Most previous studies on improving ASD performance have only focused on the use of normal data. 
In this study, we explored the use of small amounts of anomalous data for model training, based on the assumption that such data should be readily available since machines monitored by ASD systems are likely to occasionally experience anomalous states. 
In this study we have proposed a novel ASD method called Serial-OE, which combines the use of outlier exposure with a serial  method.
The proposed method improves ASD performance when using normal and pseudo-anomalous data and when utilizing a small amount of real anomalous data combined with the pseudo-anomalous data, which resulted in the best ASD performance in our evaluation experiment. 

This study also evaluated the performance of each component of the proposed method, including the use of loss functions during training, visualization, impact of machine ID information, and performance when the normal data was contaminated with anomalous data, which has provided useful insights for improving ASD performance, and for applying the proposed method under various conditions.

This study focused not only on comparing the performance of the proposed method with the performance of conventional ASD methods when using normal data from the DCASE2020 Task2 dataset, but also on the practical utilization of anomalous data to further boost ASD performance.
It may be obvious that utilizing anomalous data improves detection performance, however it is not obvious that a significant increase in performance can be achieved by using only a 10-second sample of the sound of a machine malfunctioning.
Although we are aware of the long-held view that collecting anomalous data is not practical (an assumption which we challenge), this study was nevertheless able to demonstrate the significant benefits that can be obtained by utilizing such data, while also identifying several areas with high potential for achieving further incremental increases in ASD performance. 
In order to move our proposed system into the real world, various issues must still be addressed, such as countermeasures against domain shift, application to edge devices, restrictions on collectible data, automated collection of anomalous data, and data management. 
Therefore, we will continue our research in order to develop a highly accurate ASD system that can operate in the real world.

\section*{Biographies}
\noindent\textbf{Ibuki Kuroyanagi} received the B.E. and M.E. degrees from Nagoya University in 2021 and 2023, respectively. He is currently pursuing a Ph.D. degree at the Graduate School of Informatics, Nagoya University. 
His research interests are anomalous sound detection and sound event detection. 

\vspace{1em} 
\noindent\textbf{Tomoki Hayashi} received the B.E. degree in engineering, and the M.E. and Ph.D. degrees in information science from Nagoya University, Aichi, Japan, in 2014, 2016, and 2019, respectively. He is currently a Postdoctoral Researcher with Nagoya University, Nagoya, Japan and the Chief Operating Officer of Human Dataware Lab. Co., Ltd., Nagoy. He is currently the main Developer of the end-to-end speech processing toolkit ESPnet. His research interests include statistical speech and audio signal processing. 

\vspace{1em} 
\noindent\textbf{Kazuya Takeda} received the B.E., M.E., and Ph.D. degrees from Nagoya University. He was with Advanced Telecommunication Research Laboratories and KDD Research and Development Laboratories. In 1995, he joined Nagoya University, where he started a research group for signal processing applications. He is currently a Professor with the Graduate School of Informatics and the Green Mobility Collaborative Research Center, Nagoya University. His research interest includes behavior signal processing, including driving behavior.

\vspace{1em} 
\noindent\textbf{Tomoki Toda} received the B.E. degree from Nagoya University, Nagoya, Japan, in 1999, and the M.E. and D.E. degrees from Nara Institute of Science and Technology (NAIST), Ikoma, Japan, in 2001 and 2003, respectively. He was a Research Fellow with the Japan Society for the Promotion of Science, from 2003 to 2005. He was then an Assistant Professor (2005–2011) and an Associate Professor (2011–2015) at NAIST. From 2015, he has been a Professor with the Information Technology Center, Nagoya University. His research interests include statistical approaches to speech processing.
\section*{References}
\printbibliography

\end{document}